\documentclass[12pt]{article}
\usepackage[utf8]{inputenc}
\usepackage{amsmath}
\usepackage{amssymb}
\usepackage{stmaryrd} 
\usepackage{bm}
\usepackage{cancel}
\usepackage[normalem]{ulem}
\usepackage[linesnumbered,ruled,vlined]{algorithm2e}
\usepackage{graphicx}
\usepackage{caption}
\usepackage{subcaption}
\usepackage[normalem]{ulem}
\usepackage{mathtools} 
\usepackage{authblk} 

\usepackage{tikz}
\usetikzlibrary{positioning}
\usetikzlibrary{shapes.multipart}
\usetikzlibrary{shapes.geometric}
\usetikzlibrary{arrows.meta}
\usetikzlibrary{backgrounds}
\usetikzlibrary{shapes}
\usetikzlibrary{arrows,decorations.pathmorphing}
\usetikzlibrary{positioning}
\usetikzlibrary{calc}
\usetikzlibrary{intersections}

\newcommand{\comment}[1]{}

\usepackage[sorting=none, backend=bibtex]{biblatex}
\usepackage[colorlinks=true,
	linkcolor=black,
    filecolor=magenta,
    citecolor=magenta,      
    urlcolor=cyan]{hyperref}
\makeatletter
\renewcommand*{\eqref}[1]{%
  \hyperref[{#1}]{\textup{\tagform@{\ref*{#1}}}}%
}
\usepackage{cleveref}
\makeatother
\usepackage{notoccite}
\title{Divergence error based $p$-adaptive discontinuous Galerkin solution of time-domain Maxwell's equations}
\author[,1]{Apurva Tiwari\thanks{Corresponding author. \\ \textit{Email addresses: }\href{mailto:apurva.t@aero.iitb.ac.in}{apurva.t@aero.iitb.ac.in}  (A. Tiwari), avijit@aero.iitb.ac.in (A. Chatterjee) }}
\author[1]{Avijit Chatterjee}
\affil[1]{\small \textit{Department of Aerospace Engineering, Indian Institute of Technology Bombay, India}}
\date{}

\begin{document}
\maketitle
\section{Abstract}
A $p$-adaptive discontinuous Galerkin time-domain method is developed to obtain high-order solutions to electromagnetic scattering problems. A novel feature of the proposed method is the use of divergence error to drive the $p$-adaptive method. 
The nature of divergence error is explored and that it is a direct consequence of the act of discretization is established. Its relation with relative truncation  error is formed which enables the use of divergence error as an inexpensive proxy to truncation error. Divergence error is used as an indicator to dynamically  identify and assign spatial operators of varying accuracy to substantial regions in the computational domain. This results in a reduced computational cost than a comparable discontinuous Galerkin time-domain solution using uniform degree piecewise polynomial bases throughout. \\ \\
\textit{Keywords:} Discontinuous Galerkin; Divergence error; Higher-order;  Time-domain; Maxwell's equations
\section{Introduction}
A variety of physical phenomena are modelled as systems of partial differential equations that admit divergence-free solutions. In some of these, like the incompressible Euler and Navier-Stokes equations, this condition of the solution being divergence-free is enforced explicitly. In certain other systems like the time-domain Maxwell's equations, the usual practice is to incorporate the solenoidal condition within the evolution equations, combined with the requirement that the initial conditions be solenoidal. It relies on the reasoning that if field variables are initially divergence-free, they remain so when evolved in time using the first order div-curl equations. Not all numerical schemes try to satisfy the solenoidal condition. \par
The finite difference time domain (FDTD) method proposed by Yee \cite{Yee} to solve the time-domain Maxwell's equations, satisfies the divergence-free condition by design. It uses a grid where different components of the field variables are computed at staggered spatial and temporal points. Assuming zero initial conditions, the time derivative of the net flux from the surface of a Yee cell remains a constant zero, satisfying Gauss' Law \cite{Hagness2005}. Advancements in higher order Godunov schemes for problems in computational electromagnetics (CEM) gave rise to the finite volume time domain (FVTD) \cite{ShankarFVTD} and discontinuous Galerkin time domain (DGTD) methods \cite{Hesthaven}. These schemes do not account for the divergence constraint in the time domain Maxwell system. In literature, there are various approaches to meet the constraint imposed by Gauss’ law. A divergence cleaning step is often added that solves a Poisson equation for a correction potential. Assous \cite{Assous1993} used a constrained variational formulation of Maxwell's equations and applied a penalization technique. In \cite{Munz}, Munz et. al. reformulated the constrained Maxwell's equations and introduced a coupling term into Gauss’ law, rendering a perfectly hyperbolic system of equations. This made for a natural extension of the explicit methods for Maxwell's equations to a purely hyperbolic system. In DGTD, with standard piecewise polynomial spaces used and no dedicated measures for constraint preservation taken, it is observed that global divergence errors are $p$-th order small when using polynomial bases of degree $p$ to represent the solution \cite{CockburnLDG}. DG schemes based on locally divergence-free bases \cite{CockburnLDG,Li2005} and globally divergence constraint preserving methods, for hyperbolic conservation laws \cite{Chandrashekar2019} and for ideal magnetohydrodynamics (MHD) \cite{Li2011,Li2012} have been presented in literature. 
\par Another aspect is that divergence errors accruing in conservative higher-order formulations do not significantly impact the overall accuracy of the solution \cite{Munz,Cioni} and are often disregarded in practice. In \cite{Cioni}, Cioni et. al used a mixed finite volume/finite element method to show that divergence error, despite being linked to the accuracy of the solver and the underlying discretization, does not hamper the formal accuracy of the solution. \par
In this paper, we propose another point of view. Rather than ignoring or eliminating the naturally occurring divergence error, we suggest constructively using this error in divergence, to improve spatial accuracy  in conservative, non FDTD frameworks for solving the time domain Maxwell's equations. Since solving the evolution equations does not decrease divergence error in computations, it constantly tracks truncation error. Truncation error is what many adaptive algorithms rely on, and use various methods to estimate it \cite{Kompenhans2016a, Rueda-Ramirez2019, KompenhansRubio}. We show that the relative divergence error is linked to the relative truncation error, and that it can be treated as a proxy. This makes for a readily computable indicator for adaptive algorithms, as local divergence can be computed accurately and can be included in an existing code with minimal revision. The proposed divergence based error indicator may be utilized to drive adaptive methods that assign spatial operators of varying accuracy in the computational domain, with a motive of achieving desired levels of accuracy using fewer degrees of freedom (DOFs).\par
We begin with the transverse magnetic (TM) mode of the time-domain Maxwell’s equations to formulate the formal relation between the relative truncation error and the associated relative divergence errors. We study how the relative divergence error evolves and propagates in the computational domain. Here, relative refers to the difference between quantities computed using different discretizations \cite{ChatCiCP}. We extend this definition to incorporate it in a $p$-adaptive DGTD framework. Different levels of discretizations are obtained by discrete spatial operators formed using polynomial bases of varying degrees. The formulated relation is applied on plane wave solutions and the resultant simplified expressions obtained, are numerically verified by solving canonical problems using DGTD. Numerical solutions including scattering off circular cylinders and a semi-open cavity, using a $p$-adaptive algorithm based on divergence error are presented.
\label{sec:Introduction}

\section{Discontinuous Galerkin time domain method}
This section briefly outlines the discontinuous Galerkin time domain method. A detailed account can be found in the text by Hesthaven et. al. \cite{Hesthaven}, part of which is presented in this section.
\par Consider a system of hyperbolic conservation laws,
\begin{equation}
\frac{\partial \mathbf{u}}{\partial t}+\nabla\cdot\mathcal{F} = \mathbf{s}
\label{eq:cons}
\end{equation}
where, $\mathbf{u}=\left[q_1, \cdots, q_n\right]^T$ is a vector of conserved variables,  $\mathcal{F}$ is the flux tensor depending on $q$, and $s$ is a source term.
This system is representative, among others, of the Maxwell's curl equations. For instance, consider the vector components in the 2D $TM_z$ mode, for which
\begin{equation}
\mathbf{u}=\left(
\begin{array}{c}
B_x \\ B_y \\ D_z
\end{array}
\right), \hspace*{0.3cm}
\mathbf{f}=\left(
\begin{array}{c}
0 \\ -D_z/\epsilon \\ -B_y/\mu
\end{array}
\right), \hspace*{0.3cm}
\mathbf{g}=\left(
\begin{array}{c}
D_z/\epsilon \\ 0 \\ B_x/\mu
\end{array}
\right), \hspace*{0.3cm}
\mathbf{s}=\left(
\begin{array}{c}
0 \\ 0 \\ -J_{iz}
\end{array}
\right).
\label{eq:TM}
\end{equation} 
and in the 2D $TE_z$ mode given by,
\begin{equation}
\mathbf{u}=\left(
\begin{array}{c}
B_z \\ D_x \\ D_y
\end{array}
\right), \hspace*{0.25cm}
\mathbf{f}=\left(
\begin{array}{c}
D_z/\epsilon \\ 0 \\ B_z/\mu
\end{array}
\right), \hspace*{0.25cm}
\mathbf{g}=\left(
\begin{array}{c}
-D_x/\epsilon \\ -B_z/\mu \\ 0 
\end{array}
\right), \hspace*{0.25cm}
\mathbf{s}=\left(
\begin{array}{c}
0 \\ -J_{ix} \\ -J_{iy}
\end{array}
\right).
\end{equation}
with $\mathcal{F}=\left[\mathbf{f},\mathbf{g}\right]$ as corresponding flux vectors in the $x$ and $y$ directions, along with constituent relations $\mathbf{B}=\mu\mathbf{H}, \mathbf{D}=\epsilon \mathbf{E}$.
The spatial domain $\Omega$ is triangulated as $K$ elements, $\Omega\simeq\Omega_h=\bigcup_{k=1}^K\mathsf{D}^k$ and its boundary $\partial\Omega_h$, where $\mathsf{D}^k$ is a straight-sided triangle with the triangulation assumed to be geometrically conforming.
\par The solution $\mathbf{u}$ is approximated as $\mathbf{u}_h$ and expressed locally as a polynomial expansion using local polynomial bases of order $p$, defined on element $\mathsf{D}^k$. The global solution is given by,
\begin{equation}
\mathbf{u}(\bm{x},t)\simeq \mathbf{u}_h(\bm{x},t)=\bigoplus_{k=1}^K \mathbf{u}_{h,p}^k(\bm{x},t) \in \mathsf{V}_h=\bigoplus_{k=1}^K \left\lbrace\psi_n(\mathsf{D}^k)\right\rbrace_{n=1}^{N_p}
\label{eq:uapprox}
\end{equation}
Here $\psi_n(\mathsf{D}^k)$ is a two-dimensional polynomial basis defined on element $\mathsf{D}^k$. $\mathsf{V}_h$ is a space of piecewise polynomial functions on $\Omega_h$.
It is required that the residual be orthogonal to all test functions in $\mathsf{V}_h$, and the choice that the test functions and the basis functions span the same space leads to the Galerkin scheme.
\begin{equation}
\int_\Omega\left(\frac{\partial\mathbf{u}_h}{\partial t}+\frac{\partial\mathbf{f}_h}{\partial x}+\frac{\partial\mathbf{g}_h}{\partial y}\right)\psi_n(\bm{x})=0; \hspace{0.2cm} \forall \psi_h\in\mathsf{V}_h
\label{eq:resdiualortho}
\end{equation}
Here, we have assumed $\mathbf{s}$ to be zero, \textit{i.e.} free space, for ease of explanation with minimal loss of generality. A component $u$ of the solution, is locally expressed as a $p$-th order polynomial expansion $u_{h,p}^k$,
\begin{equation}
\bm{x}\in\mathsf{D^k} : u_{h,p}^k(\bm{x},t)=\sum_{i=1}^{N_p}u_{h,p}^k(\bm{x}_i,t)\ell_i^k(\bm{x}).
\label{eq:solapprox}
\end{equation}
Here, $N_p=(p+1)(p+2)/2$, is the number of nodes in element $\mathsf{D}^k$. $\ell_i^k(\bm{x})$ is a $p$-th order multidimensional Lagrange polynomial, based on nodes $\bm{x}_i$ on $\mathsf{D}^k$. 
Note that the local order $p$ is allowed to vary across elements, \textit{i.e.} $p=p(k)$. Simple manipulation of eq. \eqref{eq:resdiualortho} leads to the local statements,
\comment{\begin{equation}
\int_{\mathsf{D}^k}\left(\frac{\partial \mathbf{u}_h^k}{\partial t}\ell_i^k(\bm{x})-
\mathcal{F}_h^k\cdot\nabla\ell_i^k(\bm{x})\right)d\bm{x}=-\int_{\partial\mathsf{D}^k}\bm{\hat{n}}\cdot\mathcal{F}^*\ell_i^k(\bm{x})d\bm{x},
\label{eq:weakform}
\end{equation} and}
\begin{equation}
\int_{\mathsf{D}^k}\left(\frac{\partial \mathbf{u}_{h,p}^k}{\partial t}+
\nabla\cdot\mathcal{F}_{h,p}^k\right)\ell_i^k(\bm{x})d\bm{x}=\int_{\partial\mathsf{D}^k}\bm{\hat{n}}\cdot\left[\mathcal{F}_{h,p}^k-\mathcal{F}^*\right]\ell_i^k(\bm{x})d\bm{x},
\label{eq:strongform}
\end{equation}
where $\bm{\hat{n}}$ is the local outward pointing normal to the element boundary $\partial \mathsf{D}^k$. Eq. \eqref{eq:strongform} is referred to as the strong form, obtained by integrating by parts  eq. \eqref{eq:resdiualortho} twice. $\mathcal{F}^*$ is the upwind flux,
\begin{equation}
\bm{\hat{n}}\cdot\left(\mathcal{F}_{h,p}^k-\mathcal{F^*}\right)=\frac{1}{2}
\begin{pmatrix}
&\hat{n}_y\left[E_z\right]+\hat{n}_x\llparenthesis\mathbf{H}\rrparenthesis-\left[H_x\right] \\
-&\hat{n}_x\left[E_z\right]+\hat{n}_y\llparenthesis\mathbf{H}\rrparenthesis-\left[H_y\right] \\
&\hat{n}_y\left[H_x\right]-\hat{n}_x\left[H_y\right]-\left[E_z\right]
\end{pmatrix},
\label{eq:upwindflux}
\end{equation}
where the following notation is used,
\begin{equation*}
\left[q\right]=q^--q^+=\bm{\hat{n}}\cdot\llparenthesis q\rrparenthesis.
\end{equation*}
Here, $q^-$ and $q^+$ are limits of the values of $q$ at the  interface from the interior and exterior of an element.
\par
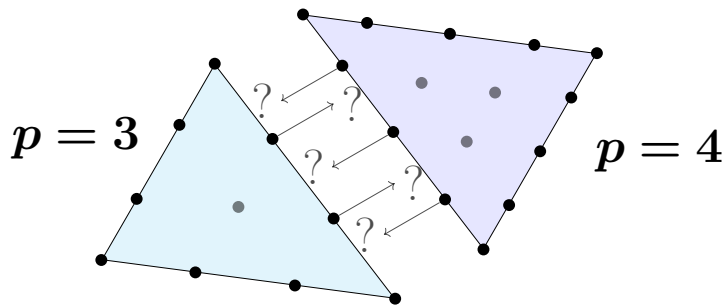
\begin{figure}
\centering
\scalebox{0.75}{
\begin{tikzpicture}
\tikzstyle{obj}  = [circle, minimum width=6pt, fill, inner sep=0pt]

\node[isosceles triangle,
    draw,
    fill=cyan!10,
    rotate=-30,
    minimum size =4cm] (Tl) at (0,0){};
    
\node[isosceles triangle,
    draw,
    fill=blue!10,
    rotate=150,
    minimum size =4cm] (Tr) at (4.5,1.5) {};

\node[obj] (tl1) [above=0cm of Tl.apex, anchor=center] {};    
\node[obj] (tl2) [above=0cm of Tl.20, anchor=center] {};
\node[obj] (tl3) [above=0cm of Tl.80, anchor=center] {};
\node[obj] (tl4) [above=0cm of Tl.left corner, anchor=center] {};
\node[obj] (tl5) [above=0cm of Tl.150, anchor=center] {};
\node[obj] (tl6) [above=0cm of Tl.210, anchor=center] {};
\node[] (p3) [above left= of tl6, anchor=center] {\huge $\bm{p=3}$};
\node[obj] (tl7) [above=0cm of Tl.right corner, anchor=center] {};
\node[obj] (tl8) [above=0cm of Tl.280, anchor=center] {};
\node[obj] (tl9) [above=0cm of Tl.340, anchor=center] {};

\path[name path=line 1] (Tl.left side) -- (tl7);
\path[name path=line 2] (tl5) -- (tl9);
\path [name intersections={of=line 1 and line 2,by=tl10}];
\node[obj, opacity=0.5] at (tl10) {};

\node[obj] (tr1) [above=0cm of Tr.apex, anchor=center] {};
\node[obj] (tr2) [above=0cm of Tr.10, anchor=center] {};
\node[obj] (tr3) [above=0cm of Tr.left side, anchor=center] {};
\node[obj] (tr4) [above=0cm of Tr.100, anchor=center] {};
\node[obj] (tr5) [above=0cm of Tr.left corner, anchor=center] {};
\node[obj] (tr6) [above=0cm of Tr.140, anchor=center] {};
\node[obj] (tr7) [above=0cm of Tr.lower side, anchor=center] {};
\node[] (p4) [right=2cm of tr7, anchor=center] {\huge $\bm{p=4}$};
\node[obj] (tr8) [above=0cm of Tr.220, anchor=center] {};
\node[obj] (tr9) [above=0cm of Tr.right corner, anchor=center] {};
\node[obj] (tr10) [above=0cm of Tr.260, anchor=center] {};
\node[obj] (tr11) [above=0cm of Tr.right side, anchor=center] {};
\node[obj] (tr12) [above=0cm of Tr.350, anchor=center] {};
\path[name path=line 1] (tr3) -- (tr11);
\path[name path=line 2] (Tr.apex) -- (Tr.lower side);
\path [name intersections={of=line 1 and line 2,by=tr13}];
\path[name path=line 1] (tr3) -- (tr7);
\path[name path=line 2] (tr5) -- (Tr.right side);
\path [name intersections={of=line 1 and line 2,by=tr14}];
\node[obj, opacity=0.5] at (tr13) {};
\node[obj, opacity=0.5] at (tr14) {};

\path[name path=line 1] (tr7) -- (tr11);
\path[name path=line 2] (Tr.left side) -- (Tr.right corner);
\path [name intersections={of=line 1 and line 2,by=tr15}];
\node[obj, opacity=0.5] at (tr15) {};

\draw[->, opacity=0.7] (tl3) -- ++(30:1.25cm) node[right] {\huge ?};
\draw[->, opacity=0.7] (tl2) -- ++(30:1.25cm) node[right] {\huge ?};
\draw[->, opacity=0.7] (tr2) -- ++(-150:1.25cm) node[left] {\huge ?};
\draw[->, opacity=0.7] (tr3) -- ++(-150:1.25cm) node[left] {\huge ?};
\draw[->, opacity=0.7] (tr4) -- ++(-150:1.25cm) node[left] {\huge ?};
\end{tikzpicture}}
\caption{``Misaligned" neighbouring nodes at common edges in a $p$-variable nodal DG framework}
\label{fig:misaligned_neighbours}
\end{figure}

 We now briefly discuss implementation considerations. Computing the numerical flux needs knowledge of the corresponding neighbours of each node, the $u^+$ as referred to in this section. To accommodate a variation of $p$ across elements, it is imperative to address situations like the one shown in fig. \ref{fig:misaligned_neighbours}. Neighbouring elements are at different levels of spatial accuracy and therefore have an unequal number of nodes representing the solution in each element. In a nodal DG setup, this leads to the problem of finding the limit of $u$ at the interface from the exterior. Here, we resort to a polynomial interpolation  \cite{Moxey} of the data using eq. \eqref{eq:solapprox}, from a neighbouring element to the nodes required. So in fig. \ref{fig:misaligned_neighbours}, interpolation of the solution in the bottom triangle, to nodes on the common edge between its neighbour is computed, which acts as the $u^+$. This operation is done for every pair of common edges. Another concern is truncation of the outer boundaries of the computational domain. For scattering problems in particular, we require that the scattered field dampens as it moves sufficiently far away from the scatterer to safely truncate the domain \cite{Hesthaven}. To achieve this, the domain is padded with a perfectly matched layer (PML) along the outer boundaries such that they do not produce spurious oscillations at their interface with the inner domain \cite{Abarbanel1998}.

\section{Divergence Error}
\subsection{Dependence of divergence on discretization}
In the absence of dedicated divergence constraint preserving  methods, numerical solutions to the time-domain Maxwell's equations using conventional finite volume/finite element methods usually show small values of divergence that can be seen more as numerical artefacts \cite{Ramshaw1983}.
In \cite{Cioni}, Cioni et. al. showed that the finite volume method used with unstructured meshes, did not conserve exactly the divergence conditions and that the divergence error thus obtained, showed a strong dependence on the underlying discretization. The method presented, was used to solve the Maxwell's equations to obtain a time-dependent scattered field when an object is illuminated by an incident plane wave. The $L_\infty$ divergence error norms reach a harmonic steady state with an average amplitude shown to be $O\left(h\right)$, $h$ being the mesh size. Spatial accuracy of the scheme is also shown to have a great influence on the divergence error. The divergence error amplitude reduced by a factor of five, comparing a first order and a quasi-third order scheme in space \cite{Cioni}. This strong dependence is also seen in the data presented in \cite{CockburnLDG}, where Cockburn et. al. developed a locally divergence-free discontinuous Galerkin method to solve the time-domain Maxwell's equations. The standard and the locally divergence-free methods were compared in terms of the discretization and divergence error norms of the solution obtained. The error norms tabulated for the case of the advection of a smooth TM wave, show that the divergence error converged at a rate of $p$, with untreated $p$-th degree polynomial bases. The proposed locally divergence-free bases resulted in a lower divergence error, but it converged at a rate similar to that obtained with the standard polynomial bases. This convergence of divergence error with discretization is verified by us although the results are not presented here, they being identical to those obtained in \cite{CockburnLDG}. Therefore, these tests suggest a strong dependence of the divergence error on discretization. \par
Despite the divergence errors, the solution itself and the resulting physical quantities of interest show expected accuracies. For example, in the same paper  \cite{CockburnLDG} by Cockburn et. al.,  it can also be observed that the discretization error norms with locally divergence-free bases, do not differ significantly from the ones obtained using bases prone to divergence error. This again is verified by us without repeating the results here.  Following up with \cite{Cioni}, Cioni et. al. used a mixed finite volume/finite element method, not accounting for the divergence constraints explicitly. They observed that the divergence errors did not distort the radar cross section(RCS) obtained and concluded that the divergence error did not have a significant influence on the higher order accurate solution. To summarize, these observations suggests that divergence error in higher order conservative methods does show a dependence on the discretization but has an insignificant influence on the formal accuracy of the solution, especially for higher order methods. Hence, such divergence errors can often be safely ignored. The exception to this is only when the motivation to eliminate divergence error comes rather from the physics to be captured in higher order solutions of the time domain Maxwell's equations. In magnetohydrodynamics for instance, Brackbill and Barnes \cite{Brackbill1980} showed that nonzero numerical divergence causes unphysical fluid velocities, parallel to the magnetic field vector. Also, in applications such as those in magnetohydrodynamics, it is regular practice to ensure that the divergence constraints are met \cite{Altmann2009,Wheatley2010,Alvarez2021}. 

\subsection{Relation between relative divergence and relative truncation errors}
\label{sec:causalreln}
We observe that the Gauss' law relations, $\nabla\cdot \mathbf{B}=0$ and $\nabla\cdot \mathbf{D}=0$ (in free space), are not part of the evolution \cref{eq:cons,eq:TM}. According to the evolution equations in their continuous form, divergence of the field variables $\mathbf{B}$ and $\mathbf{E}$ is time independent. Therefore, if initially this divergence is zero, it will remain zero. Thus, the Gauss' law is not a dynamical condition on variables $\mathbf{B}$ and $\mathbf{E}$, rather a constraint on the initial conditions. Therefore, the fact that the Gauss' law is satisfied in the continuous case, is a consequence of the evolution equations \cref{eq:cons,eq:TM}, coupled with appropriate initial conditions \cite{Ramshaw1983}. While formulating numerical methods to solve the time domain Maxwell's equations, satisfying the divergence constraint is theoretically shown to be a consequence of the curl equations and are considered redundant, if satisfied initially \cite{Cioni, Taflove}. Thus, with numerical methods approximating only the evolution equations, the differential property that the divergence is rigorously independent of time, is generally not preserved \cite{Ramshaw1983}. With numerical tests showing a dependence of divergence error on the discretization, in this section we establish it formally.
\par Consider a subsystem of the $TM_z$ eq. \eqref{eq:cons} with \eqref{eq:TM}, consisting of only in-plane components $B_x$ and $B_y$, such that
\begin{equation}
\mathbf{U}=
\begin{pmatrix}
B_x \\ B_y 
\end{pmatrix}
=\mathbf{B}, \hspace*{0.3cm}
\mathbf{F}=
\begin{pmatrix}
0 \\ -D_z/\epsilon 
\end{pmatrix}\hat{i}
, \hspace*{0.3cm}
\mathbf{G}=
\begin{pmatrix}
D_z/\epsilon \\ 0 
\end{pmatrix}\hat{j}
, \hspace*{0.3cm}
\label{eq:subsys}
\end{equation}
with $\mathbf{S}=0$ for simplicity, and $\hat{i}, \hat{j}$ being unit vectors along the $+x$ and $+y$ directions respectively. Conservation form for this subsystem came then be written in Cartesian co-ordinates as,
\begin{equation}
\frac{\partial\mathbf{U}}{\partial t}+\nabla\cdot\mathcal{F} = \frac{\partial\mathbf{U}}{\partial t}+\frac{\partial\mathbf{F}}{\partial x}+\frac{\partial\mathbf{G}}{\partial y}=0,
\label{eq:consforsub}
\end{equation}
where the flux is given as
\begin{equation*}
\mathcal{F}=[\mathbf{F},\mathbf{G}].
\end{equation*}
Writing the vector equation eq. \eqref{eq:consforsub} component-wise,
\begin{equation}
\frac{\partial}{\partial t}\begin{pmatrix}
B_x \\ B_y
\end{pmatrix}
+\frac{\partial}{\partial x}\begin{pmatrix}
0 \\ -D_z/\epsilon
\end{pmatrix}
+\frac{\partial}{\partial y}\begin{pmatrix}
D_z/\epsilon \\0
\end{pmatrix}=0.
\end{equation}
\begin{equation}
\frac{\partial}{\partial t}\begin{pmatrix}
B_x \\ B_y
\end{pmatrix}
+\begin{pmatrix}
\frac{\partial D_z/\epsilon}{\partial y} \\ \\
          -\frac{\partial D_z/\epsilon}{\partial x}
\end{pmatrix}=0.
\label{eq:continuousvector}
\end{equation}
A representation of eq. \eqref{eq:continuousvector} as a time-dependent PDE, is given by
\begin{equation}
\frac{\partial \mathbf{U}}{\partial t} = \mathbf{R}\left(\mathbf{U}\right),
\label{eq:abstractcontinuous}
\end{equation}
where $\mathbf{R}$ is the spatial operator. Consider a higher order discretization of eq. \eqref{eq:consforsub} of the form \cite{KompenhansRubio},
\begin{equation}
\int_\Omega \nabla\cdot\mathcal{F}\left(\mathbf{U}\right)\psi d\mathbf{x}=0 \rightarrow \sum_{\mathbb{D}^k\in \Omega_h}\mathbf{R}_p\left(\mathbf{U}_p\right)=0,
\end{equation}
where $\psi$ is the test function and $p$ is the degree of the polynomial bases used to represent the discrete solution $\mathbf
{U}_p$. $\mathbf{R}_p$ now represents the discrete spatial partial differential operator. The semi-discrete explicit time marching scheme thus obtained (the subscript $h$, characteristic size of the mesh is treated constant and hence dropped, for clarity) for eq. \eqref{eq:abstractcontinuous}, is given in a conventional residual form \cite{ChatCiCP} as,
\begin{equation}
\frac{d\mathbf{U}_{p}}{dt}=\mathbf{R}_{p}\left(\mathbf{U}_p\right).
\label{eq:semidiscrete_p}
\end{equation}
Another approximation on the same mesh can be formed by projecting the $p$-th order accurate solution to a space of $q$-th order basis functions,
\begin{equation}
\tilde{\mathbf{U}}_p^q=I_p^q\mathbf{U}_p,
\end{equation}
where $I_p^q$ is the transfer operator with $p>q$, and $\left(\tilde{\cdot}\right)$ represents a restricted higher order (here, $p$-th order) quantity. The semi-discrete equation for a purely $q$-th order accurate approximation  is thus,
\begin{equation}
\frac{d\mathbf{U}_{q}}{dt}=\mathbf{R}_{q}\left(I_p^q\mathbf{U}_p\right).
\label{eq:semidiscrete_q}
\end{equation} 
A relative truncation error \cite{Fraysse2012, Brandt1977} between levels $p$ and $q$,  is usually introduced as,
\begin{equation}
\bm{\tau}_p^q=I_p^q\mathbf{R}_p\left(\mathbf{U}_p\right)-\mathbf{R}_q\left(I_p^q\mathbf{U}_p\right)= \llbracket \mathbf{R}\left(\mathbf{U}\right)\rrbracket_p^q.
\label{eq:definition_tau}
\end{equation}
Here, the notation $\llbracket \left(\cdot\right) \rrbracket_p^q$ represents the difference between quantities at levels $p$ and $q$ as shown, with operator $\mathbf{R}$ of varying spatial accuracies acting on pure and restricted solution vectors $\mathbf{U}$. This relative truncation error has been used in \cite{ChatCiCP}, as a forcing function for maintaining higher ($p$-th) order solution while operating at a lower ($q$-th) order. This was proposed in the context of developing a multilevel technique in explicit time marching schemes. Thus, the semi-discrete form to be evaluated at the $q$-th level, while maintaining $p$-th order accuracy, is given as,
\begin{equation}
\frac{d\tilde{\mathbf{U}}_p^q}{dt}=\mathbf{R}_q\left(I_p^q\mathbf{U}_p\right)+\bm{\tau}_p^q.
\label{eq:semidiscrete_qt}
\end{equation}
Taking divergence of eq. \eqref{eq:semidiscrete_qt},
\begin{equation}
\frac{d\bm{\nabla}\cdot\tilde{\mathbf{U}}_q^p}{dt}=\bm{\nabla}\cdot\mathbf{R}_q\left(I_p^q\mathbf{U}_p\right)+\bm{\nabla}\cdot\bm{\tau}_p^q.
\label{eq:div_semidiscrete_qt}
\end{equation}
With the forcing provided by the relative truncation error $\bm{\tau}_p^q$, divergence of the restricted solution $\tilde{\mathbf{U}}_p^q$ is the divergence of the higher order solution $\mathbf{U}_p$, restricted to the $q$-th level.
Taking divergence of eq. \eqref{eq:semidiscrete_q},
\begin{equation}
\frac{d\bm{\nabla}\cdot\mathbf{U}_q}{dt}=\bm{\nabla}\cdot\mathbf{R}_q\left(I_p^q\mathbf{U}_p\right).
\label{eq:div_semidiscrete_q}
\end{equation}
Difference between eqs. \eqref{eq:div_semidiscrete_qt} and \eqref{eq:div_semidiscrete_q} yields,
\begin{equation}
\frac{d\left(\bm{\nabla}\cdot\tilde{\mathbf{U}}_p^q-\bm{\nabla}\cdot \mathbf{U}_{q}\right)}{dt}=\bm{\nabla}\cdot \bm{\tau}_p^q.
\label{eq:penulreln}
\end{equation}
Therefore, time derivative of the difference between divergences of the restricted and the lower order solutions is given by the divergence of the relative truncation error. $\mathbf{U}$ represents the magnetic field vector $\mathbf{B}$ in the TM mode of the time-domain Maxwell equations. 
The left hand side of eq. \eqref{eq:penulreln} is the time derivative of a  relative divergence error $\gamma_p^q$, defined here as difference between divergences of the restricted higher order and the lower order accurate solutions, $\gamma_p^q=\bm{\nabla}\cdot\tilde{\mathbf{B}}_p^q-\bm{\nabla}\cdot \mathbf{B}_{q}$. With the relative truncation and relative divergence errors defined, eq. \eqref{eq:penulreln} yields a causal relation, which in continuous form is given as,
\begin{equation}
\boxed{\frac{d\gamma_p^q}{dt}=\bm{\nabla}\cdot\bm{\tau}_p^q}.
\label{eq:causal}
\end{equation}
Eq. \eqref{eq:causal}, hereby also called the Divergence Error Evolution Equation (DEEE), represents compactly a fundamental statement that the evolution of the relative divergence error is fed by the relative truncation error acting as a source.  If $p$ is taken to be infinity, \textit{i.e.} corresponding to the continuous case, then eq. \eqref{eq:causal} suggests that it is due to the act of discretization that a divergence error is generated. The causal relation eq. \eqref{eq:causal} between the two errors, forms the rationale behind treating divergence error as a surrogate to truncation error. Note that the divergence operators used in the definition of $\gamma_p^q$ are assumed to be infinitely accurate. This is to emphasize the fact that the divergence error is contained in the discrete solution, and is not a result of the accuracy of the divergence operator used in computing $\gamma_p^q$.
\par A relevant simplification of the DEEE in the case of a plane wave, yields further insight. Consider a plane wave solution to eq. \eqref{eq:TM}, restricted to a single frequency  $\omega$ here, as a representative instance,
\begin{subequations}
\label{eq:planewavesoln}
\begin{gather}
D_z= D_{z0}e^{(\mathbf{k}\cdot\bm{x}-\omega t)}, \\
B_y=-\sqrt{\frac{\mu}{\epsilon}}D_z cos\phi, \\
B_x=\sqrt{\frac{\mu}{\epsilon}}D_z sin\phi,
\end{gather}
\end{subequations}
where $\bm{x}$ is a position vector in the $xy$ plane; $\mathbf{k}=\frac{\omega}{c}\left[\cos\phi, \sin\phi \right]^T$ is the wavenumber with $\phi$ being the angle made with the $+x$ axis.  $c=\frac{1}{\sqrt{\mu\epsilon}}$ is the speed of propagation in the medium. Again treating $\mathbf{U}$ to be representing the magnetic field vector $\mathbf{B}$, in the TM mode plane wave solution eq. \eqref{eq:planewavesoln}, and from eqs. \eqref{eq:continuousvector} and \eqref{eq:abstractcontinuous}, the flux residual $\mathbf{R}_p$ is given by,
\begin{equation}
\mathbf{R}_p=\begin{pmatrix}
           -\frac{\partial D_z/\epsilon}{\partial y} \\ \\
           \frac{\partial D_z/\epsilon}{\partial x} \\
		   \end{pmatrix}_p                               
		   =\begin{pmatrix}
		   \frac{\partial B_y}{\partial y} \frac{c}{cos\phi} \\ \\
   		   \frac{\partial B_x}{\partial x} \frac{c}{sin\phi}
		   \end{pmatrix}_p,
\end{equation}
resulting from a higher $p$-th order discretization. As in the derivation of the DEEE, we consider another discretization resulting from restriction of the higher $p$-th order solution to a lower $q$-th order with the corresponding flux residual given as $\mathbf{R}_q\left(\tilde{\mathbf{U}}_p^q\right)$. The relative truncation error $\bm{\tau}_p^q$ for the plane wave solution eq. \eqref{eq:planewavesoln} is therefore given by
\begin{equation}
\bm{\tau}_p^q= \llbracket\mathbf{R}\left(\mathbf{U}\right) \rrbracket_p^q= \begin{pmatrix}
		   \left\llbracket\frac{\partial {B_y}}{\partial y}\right\rrbracket_p^q \frac{c}{cos\phi} \\ \\
   		   \left\llbracket\frac{\partial B_x}{\partial x}\right\rrbracket_p^q \frac{c}{sin\phi}
   		   \end{pmatrix}.
\label{eq:taupq}
\end{equation}
The relative divergence error as defined earlier, represents the inherent error arising out of not numerically meeting the divergence constraint in the time-domain Maxwell's equations. Although the definition remains agnostic to the way the divergence operator is defined computationally, practically computing $\gamma$ would invariably use a discrete divergence operator. The errors associated with such an operator would also contribute to the definition of relative divergence error, now a computable discrete quantity. We now define a fully discrete relative divergence error as,
\begin{equation}
\hat{\gamma}_p^q =I_p^q\bm{\nabla}_p\cdot\mathbf{U}_p-\bm{\nabla}_q\cdot\left(I_p^q\mathbf{U}_p\right)
=\llbracket \bm{\nabla}\cdot\mathbf{U} \rrbracket_p^q,
\label{eq:definition_gammahat}
\end{equation}
which uses discrete divergence operators and includes the errors associated with them. The definition eq. \eqref{eq:definition_gammahat} is conceptually analogous to the relative truncation error defined in eq. \eqref{eq:definition_tau}, is practically computable, and relevant from an implementation point of view. Since divergence consists only of a combination of spatial derivatives of the solution vector, in the context of DGTD, the local divergence error can be computed simply using the differentiation matrices \cite{Hesthaven} that transform point values $\mathbf{U}\left(\mathbf{x}_i\right)$ to the derivatives at those points, $\mathbf{U}_h'=\mathcal{D}_r\mathbf{U}_h$, its entries defined as,
\begin{equation}
\left.\mathcal{D}_r=\frac{d\ell_j}{dr} \right\vert_{r_i}.
\end{equation}
$\hat{\gamma}_p^q$ is given component-wise by,
\begin{equation}
\hat{\gamma}_p^q = \llbracket \bm{\nabla}\cdot\mathbf{B} \rrbracket_p^q
= \left\llbracket\frac{\partial B_x}{\partial x}\right\rrbracket_p^q
+ \left\llbracket\frac{\partial {B_y}}{\partial y}\right\rrbracket_p^q.
\label{eq:gammapq}
\end{equation}
Eqs. \eqref{eq:taupq} and \eqref{eq:gammapq} yield the following relation between the fully discrete relative divergence and relative truncation errors,
\begin{equation}
\boxed{\hat{\gamma}_p^q =\bm{\tau}_p^q\cdot\frac{\bm{\phi}}{c}},
\label{eq:scalarproxy}
\end{equation}
where $\bm{\phi}=\left[\cos\phi, \sin\phi \right]^T$, is a unit vector along the direction of wave travel.
This simple relation institutes $\hat{\gamma}$ as a scalar proxy of $\bm{\tau}$.
The discrete relative divergence error is a scaled projection of the relative truncation error along the direction of travel. This is the essential link between the feature divergence, and the truncation error which facilitates its use as an indicator of truncation error.  Also, since the divergence of the solution is readily available, it serves as an inexpensive driver in adaptive algorithms. A divergence error based method would naturally seem to belong to the feature-based class of adaptive algorithms, but these mutual relations make this method, essentially a truncation error-based method. The truncation and discretization error are related to each other by the Discretization Error Transport Equation (DETE), which shows that truncation error acts as a local source of discretization error \cite{KompenhansRubio}. This made the case for preferring truncation error as a sensor for adaptive algorithms. Hence, the extensive work in the area of truncation error estimation \cite{Rueda-Ramirez2019,KompenhansRubio,Syrakos2012}. A divergence error based sensor need not rely on estimation procedures and can be accurately computed. Also, since computing divergence only uses derivatives of the solution vector, requisite routines are usually already available in an existing code, making it practically feasible to incorporate such an indicator. The cost of computation of an error indicator, and accommodating such computations in an existing code structure is what limits the use of adaptive algorithms in commerical software \cite{Zienkiewicz1987}. Also, divergence error possesses other attributes like being a scalar and co-ordinate independent, which is advantageous over keeping track of vector fields \cite{Hernandez1997}.

\section{Algorithm}
With relative divergence error identified as proxy to relative truncation error, in this section we develop a $p$ adaptive algorithm based on divergence. This requires mapping the set of local values of divergence $\nabla\cdot \mathbf{H}_{h,p}^{k}$ in every element $\mathsf{D}^k$, to a proposed order $p_{new}^k$. Note that divergence of the magnetic field vector has been taken in the TM case. Equivalently, divergence of the electric field is to be considered in the TE case. The strategy is to simply create a logarithmic mapping between the magnitudes of the divergence, and integers in the range $\left[1, N \right]$. The parameter $N$ is a predefined maximum degree of the polynomial basis functions.
\begin{enumerate}
\item The divergence resulting from a discretization shows a wide spectrum of spatial scales. The input is a set of values of the divergence in an element $k$, at a given time level $(n)$ for a discretization $(h,p)$. The objective is to come up with a new discretization $(h,p_{new})$.
\begin{equation}
div=\left\lbrace\nabla\cdot \mathbf{H}_{h,p}\left(\bm{x}_i,t^{\left(n\right)}\right)\vert_{\mathsf{D}^k}; i \in \left[1,\cdots,N_p\right], k \in \left[1, \cdots, K\right] \right\rbrace; 
\end{equation}
$div$ is a set of all local values of divergence of $\mathbf{H}$. The number of nodes $N_p$ and hence number of values of divergence, in an element $\mathsf{D}^k$, depends on the order $p$ of the local polynomial basis functions. So, the above set is $\sum_{k=1}^K N_p(p^k)$ big, referred to as \textit{nDOF} in the input to algorithm \ref{algo:logmap}.
\item Taking a logarithm of the absolute values of $div$ segregates the spatial scales. This set is then shifted and scaled by a factor, to the range $\left[0,N\right]$.
\begin{equation}
factor: \log_{10}|div|\rightarrow ord \in \left[0,N\right] 
\end{equation}
$ord$ is an intermediate data set, \textit{nDOF} in size.

\item At this stage, entries of $f$ are real numbers. Corresponding to every element $\mathsf{D}^k$, there are $N_p$ entries in $f$, indexed by $i\in\left[1,\cdots,N_p\right]$. We take the ceiling values of $f$ and since there has to be a unique $p_{new}^k$, we pick the maximum out of these. 
\begin{equation}
p_{new}^k=\max_{i\in \left[1, \cdots, N_p\right]}\left\lceil f^k_i \right\rceil
\end{equation}
This seems a conservative choice; the rationale behind it is that it is tracking the truncation error closely. In other words, out of all the nodes in an element demanding various $p$, the order of the cell, \textit{i.e.} the one applicable on all nodes should be the maximum asked for.
\item Finally, as a check, cells with all nodes having  $f=0$, are kept at the $p=1$ level. Also, cells forming the PML boundary if employed, are kept at the $p=N$ level to avoid any spurious interactions.
\end{enumerate}

The steps involved are stated concisely in algorithm \ref{algo:logmap}.

\begin{algorithm}
\SetAlgoLined
\KwIn{div = $\nabla\cdot \mathbf{H}_{h,p}^{(n)}$ or $\nabla\cdot \mathbf{E}_{h,p}^{(n)}$  \hspace*{0.1cm}\textit{shape}: $[$nDOF, 1$]$}
\KwOut{$p_{new}$ \hspace*{3.1cm}\textit{shape}: $[$K, 1$]$}
logdiv = log$_{10}$(abs(div));\\
shift = -min(logdiv);\\
scale = 1/max(logdiv+shift);\\
logdiv = (logdiv+shift)*scale; \hspace*{0.6cm} \textit{shift and scale to range $[0,1]$}\\
ord = $\lceil$N*logdiv$\rceil$; \hspace*{2.4cm} \textit{scale to $[0,N]$, take ceiling values}\\
i1=0; i2=0;\hspace*{3.55cm} \textit{to index into} ord\\
 \For{$k = [1,\cdots,K]$}{
 i1=i2+1; i2=i2+$N_p(p^k)$ ;\\
  $p_{new}(k)$=max(ord(i1:i2)); \hspace*{0.8cm} \textit{take max out of $N_p(p^k)$  values}
  
 }
 sanity check: $p_{new}^k=1$; \hspace*{1.75cm} $\forall \hspace*{0.1cm}  k$ with $p_{new}^k=0$ \\
 boundary/PML cells at $p=N$; \hspace*{0.5cm} $p_{new}\in\{1,\cdots,N\}$
 \caption{$p$-adaptation routine, logarithmic map}
\label{algo:logmap}
\end{algorithm}

\section{Results}
\subsection{Scattering off a circular cylinder} 
\label{problem:cylscat}
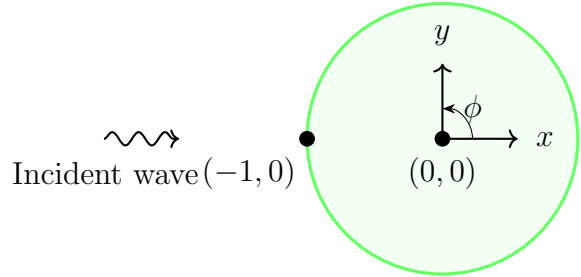
\begin{figure}
\centering
\begin{tikzpicture}[
photon/.style={decorate, decoration={snake}, draw=black},
squaregreennode/.style={circle, draw=black, fill=black, very thick, scale=0.5},
roundnode/.style={circle, draw=green!60, fill=green!5, very thick, minimum size=36mm},
roundnode2/.style={draw,shape=circle,fill=black,scale=0.5},
coord/.style={draw,shape=circle,fill=black,scale=0}
]

\node[roundnode]      (scat)         {};
\node[roundnode2,
label={[label distance=1.0cm]0:$x$},
label={[label distance=1.0cm]90:$y$},
label={[label distance=0.0cm]270:$\left(0,0\right)$},
label={[label distance=3cm]183:Incident wave},draw]      (orig) {};
\node[roundnode2,label={267:$\left(-1,0\right)$}] (leftorig) at (-1.8,0) {};
\node[] (phi) at (0.4,0.4) {$\phi$};

\draw[black, thick,->] (orig) -- (1,0);
\draw[thick,->,photon] (-4.5,0) -- (-3.5,0) ;
\draw[black, thick,->] (orig) -- (0,1);
\draw[thin, -Stealth] (.4,0) arc  (0:90:0.4);

\end{tikzpicture}\\
\caption{Schematic for the circular cylinder scattering problem}
\label{fig:schematic}
\end{figure}
The proposed algorithm is applied on the cylinder scattering problem for validation. A schematic of the problem is shown in fig. \ref{fig:schematic}. Variations of the problem in terms of electrical sizes and incident TM and TE illumination are presented.  

\begin{figure}
\centering
\includegraphics[width=\textwidth]{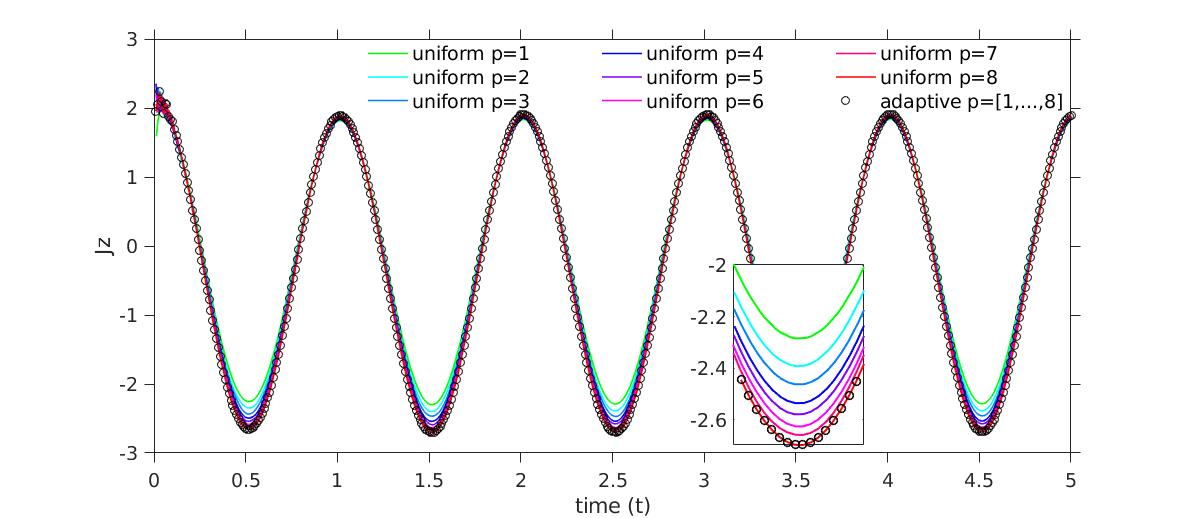}
\caption{Surface current at (-1,0) for various $p$, circular cylinder scattering}
\label{fig:Jzt_compare}
\end{figure}

The surface current at point $(-1,0)$ as shown in the schematic fig. \ref{fig:schematic}, for a cylinder 2$\lambda$ in diameter under TM illumination, is plotted in fig. \ref{fig:Jzt_compare} for various $p=1,\cdots,8$. Another instance with an adaptive routine running a combination of all levels $p=[1,\cdots,8]$ according to algorithm \ref{algo:logmap} is compared. The surface current is sinusoidal, so approximations accurate up to various orders are visibly distinguishable and suggest that a variation in $p$ does show up in the results up to at least $p=8$. Therefore, this is used as a suitable test case for comparison between the $p$-adaptive and the standard uniform $p$ methods. 

\begin{figure}
\centering
\includegraphics[width=0.75\textwidth]{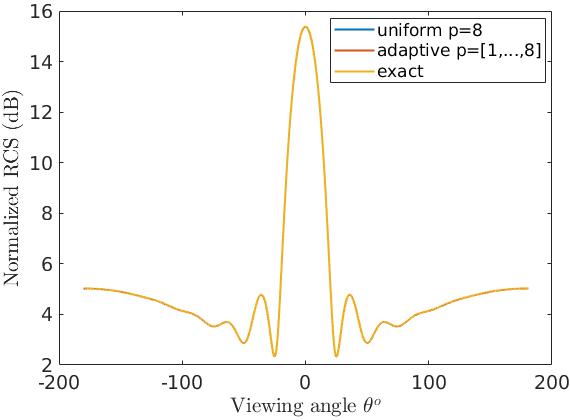}
\caption{RCS, circular cylinder scattering,  size $2\lambda$, TM illumination}
\label{fig:rcs2TM}
\end{figure}

\begin{figure}
\centering
\includegraphics[width=0.75\textwidth]{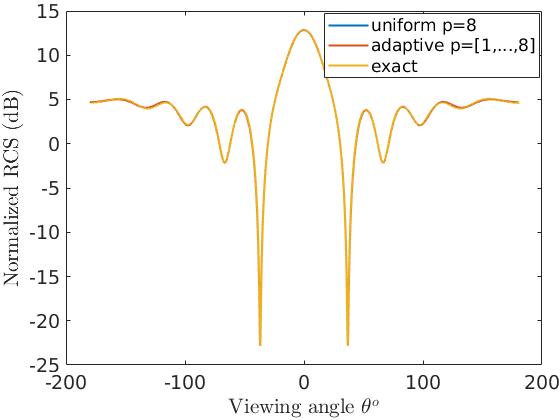}
\caption{RCS, circular cylinder scattering,  size $2\lambda$, TE illumination}
\label{fig:rcs2TE}
\end{figure}

Figs. \ref{fig:rcs2TM} and \ref{fig:rcs2TE} show the scattering width obtained, compared to the exact solutions under TM and TE illuminations respectively, for a scatterer $2\lambda$ in size.   Viewing angle $0^0$ is measured from the $+x$ axis as shown in the schematic fig. \ref{fig:schematic}. Here, the adaptive algorithm takes $N=8$, \textit{i.e}, $p=[1,\cdots,8]$. The adaptation takes place every iteration, throughout the duration of the problem including the initial unsteady phase, followed by its evolution into a harmonically steady state.

\begin{figure}
\centering
\includegraphics[width=0.75\textwidth]{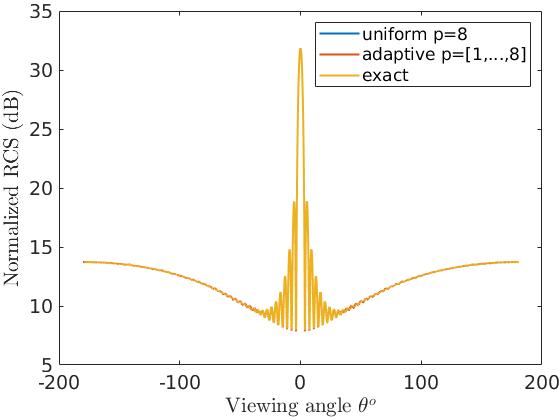}
\caption{RCS, circular cylinder scattering,  size $15\lambda$, TM illumination}
\label{fig:rcs15TM}
\end{figure}

\begin{figure}
\centering
\includegraphics[width=0.6\textwidth]{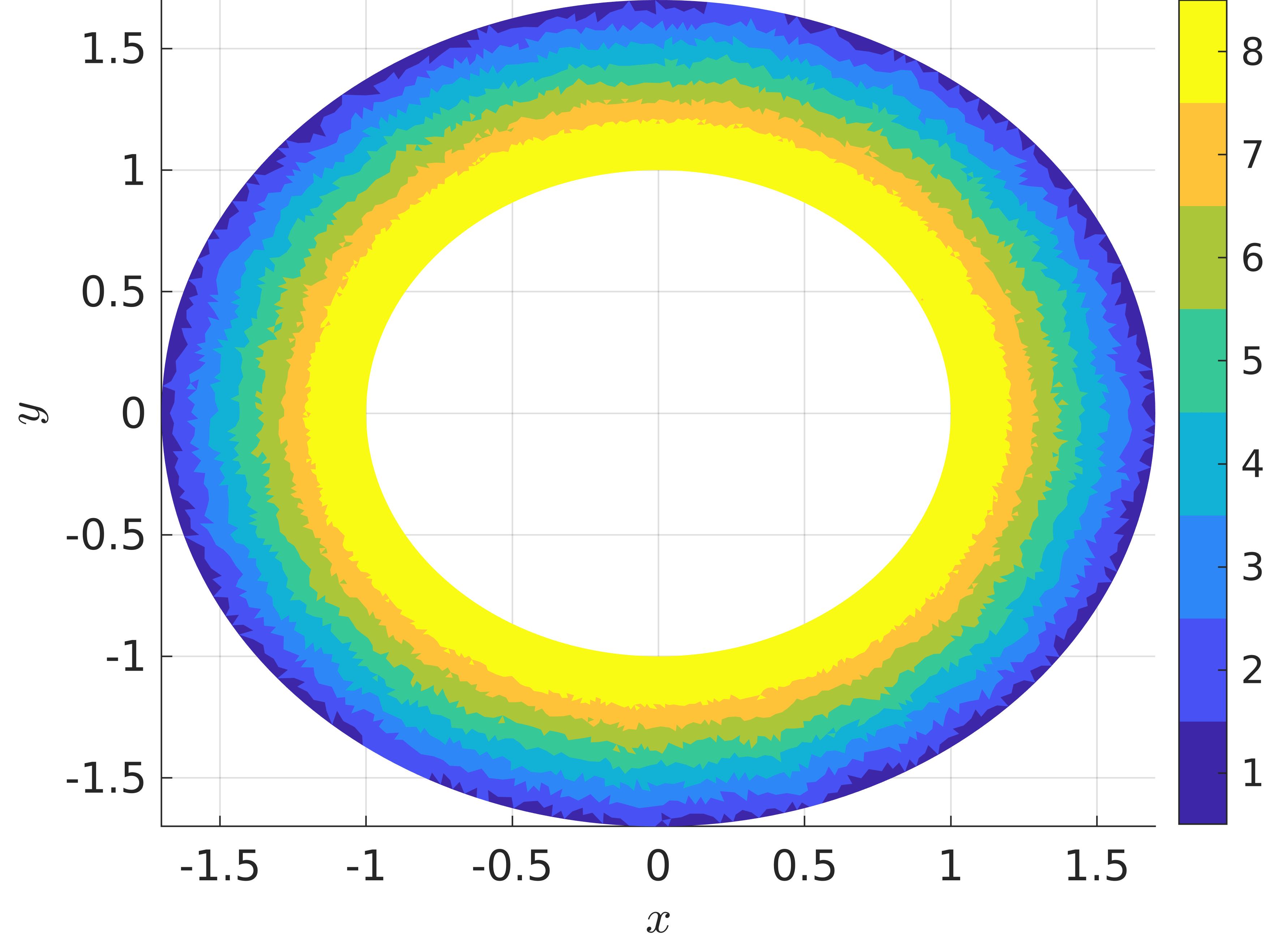}
\caption{Distribution of $p$, circular cylinder scattering,  size $15\lambda$}
\label{fig:pdis_cylscat}
\end{figure}

A similar comparison for a larger scatterer of size $15\lambda$ is shown in fig. \ref{fig:rcs15TM}. As a representative illustration, fig. \ref{fig:pdis_cylscat} shows the distribution of $p$ in the domain at the one period mark. Cells forming the scattering surface and in close proximity  are at $p=8$ and it drops radially outwards. A linear variation of $p$ in the radial direction can be seen, in accordance with algorithm \ref{algo:logmap} that created a linear mapping between the order of magnitudes of local divergence of the solution, and integers $[1,\cdots,N]$.

\subsection{Semi-open cavity}
\label{problem:SOcavity}
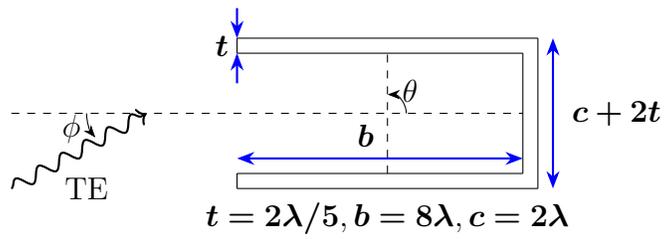
\begin{figure}
\centering
\scalebox{1}{
\begin{tikzpicture}[
photon/.style={decorate, decoration={snake}, draw=black}]
\draw[] (-2,-1) -- (2,-1);
\draw[thick, {Stealth}-{Stealth},blue] (-2,-0.6) --node [pos=0.45,above,black] {$\bm{b}$} (1.8,-0.6);
\draw[] (-2,1) -- (2,1);
\draw[] (-2,-0.8) -- (1.8,-0.8);
\draw[] (-2,0.8) -- (1.8,0.8);
\draw[] (-2,-1) -- (-2,-0.8);
\draw[] (-2,0.8) -- (-2,1);
\draw[] (2,-1) -- (2,1);
\draw[thick, {Stealth}-{Stealth},blue] (2.2,-1) --node [midway,right=0.1cm,black] {$\bm{c+2t}$} (2.2,1);
\draw[] (1.8,-0.8) -- (1.8,0.8);
\draw[thick, -{Stealth},blue] (-2,1.4) -- (-2,1);
\draw[thick, -{Stealth},blue] (-2,0.4) -- (-2,0.8);
\node[] (t) at (-2.2,0.9) {$\bm{t}$};
\draw[thin, dashed] (-5,0) -- (1.8,0);
\draw[thin, dashed] (0,-0.8) -- (0,0.8);
\draw[thick,->,photon] (-5,-1) -- (-3.2,0);
\node[] (phi) at (-4.2,-0.2) {$\phi$};
\draw[thin, -Stealth] (-4,0) arc  (180:220:0.5);
\node[] (TE) at (-4.0,-1.0) {TE};
\draw[thin, -Stealth] (0.25,0) arc  (0:90:0.25);
\node[] (theta) at (0.3,0.3) {$\theta$};
\node[] (measurements) at (0,-1.4) {$\bm{t=2\lambda/5}, \bm{b=8\lambda}, \bm{c=2\lambda}$};
\end{tikzpicture}}
\caption{Schematic for the semi-open cavity problem}
\label{fig:schematic_SOcavity}
\end{figure}

\begin{figure}
\centering
\includegraphics[width=\textwidth]{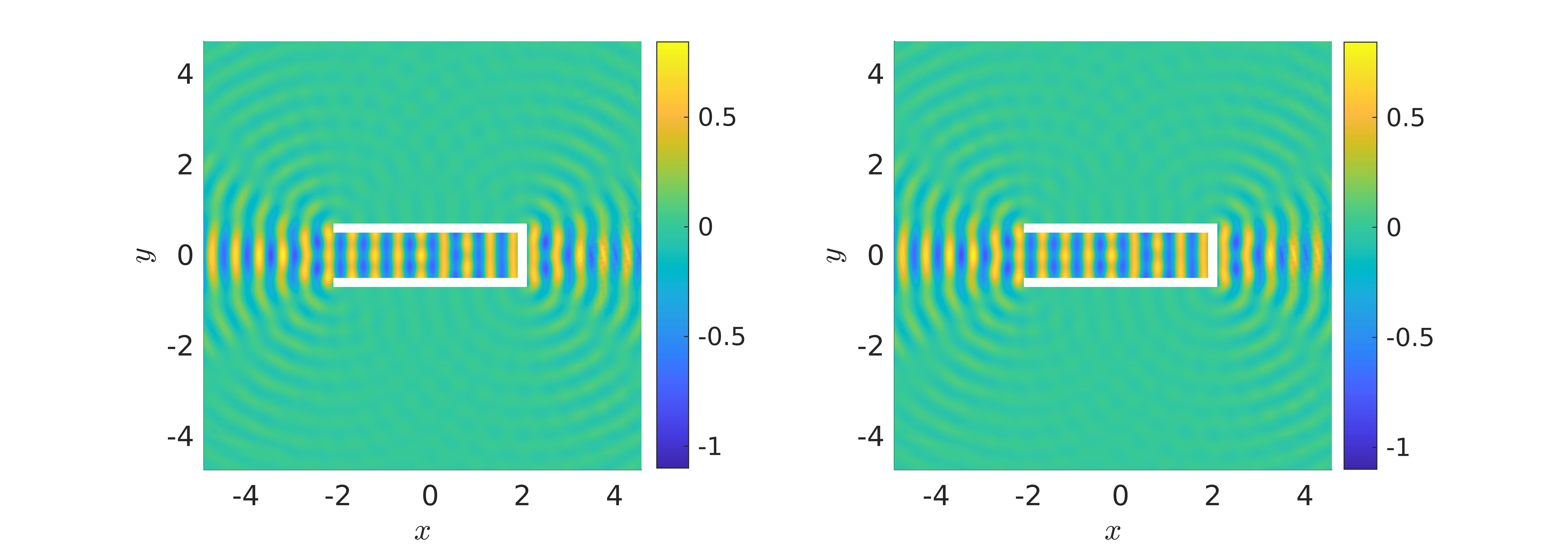}
\caption{Scattered $H_z$: uniform $p=8$ vs. adaptive $p=[1,\cdots,8]$}
\label{fig:contourHz_phi0}
\end{figure}

The adaptive algorithm is next applied to scattering from a semi-open cavity. The complexity of this problem over the cylinder as a scatterer  lies in the multiple internal reflections that the solution goes through in the cavity. Fig. \ref{fig:schematic_SOcavity} shows the schematic of the problem. A TE wave, incident at an angle $\phi$ to the $x$-axis impinges on the scatterer as shown. We present results for two such angles $\phi=0^0, 30^0$. The computational domain is padded with a PML 1$\lambda$ wide to truncate the domain. To aid visualization of the solution, fig. \ref{fig:contourHz_phi0} shows $H_z$ under TE illumination at $\phi=0^0$. The figure compares solutions obtained from the uniform $p=8$ and the adaptive $p=[1,\cdots,8]$ methods.

\begin{figure}
\centering
\includegraphics[width=0.75\textwidth]{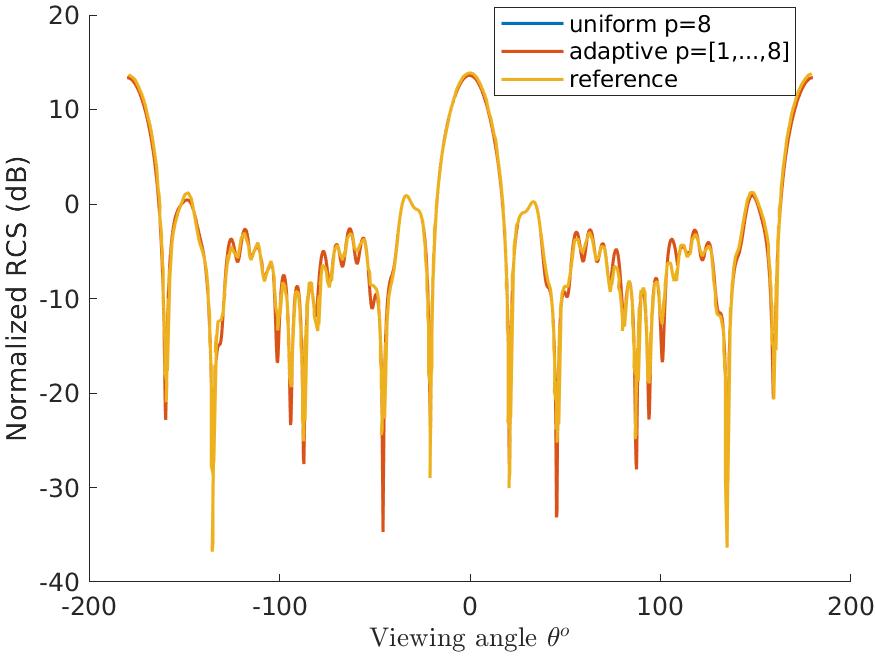}
\caption{RCS, semi-open cavity $8\lambda \times 2\lambda$, $\phi=0^0$}
\label{fig:rcsSOcavity_phi0}
\end{figure}

The RCS plots are also compared as in fig. \ref{fig:rcsSOcavity_phi0}. A 5th order accurate hybrid Galerkin solution to this problem, presented in \cite{Davies2009} has been used as reference. Results shown have been computed at the 40 period mark, the solution being well converged.

\begin{figure}
\centering
\includegraphics[width=\textwidth]{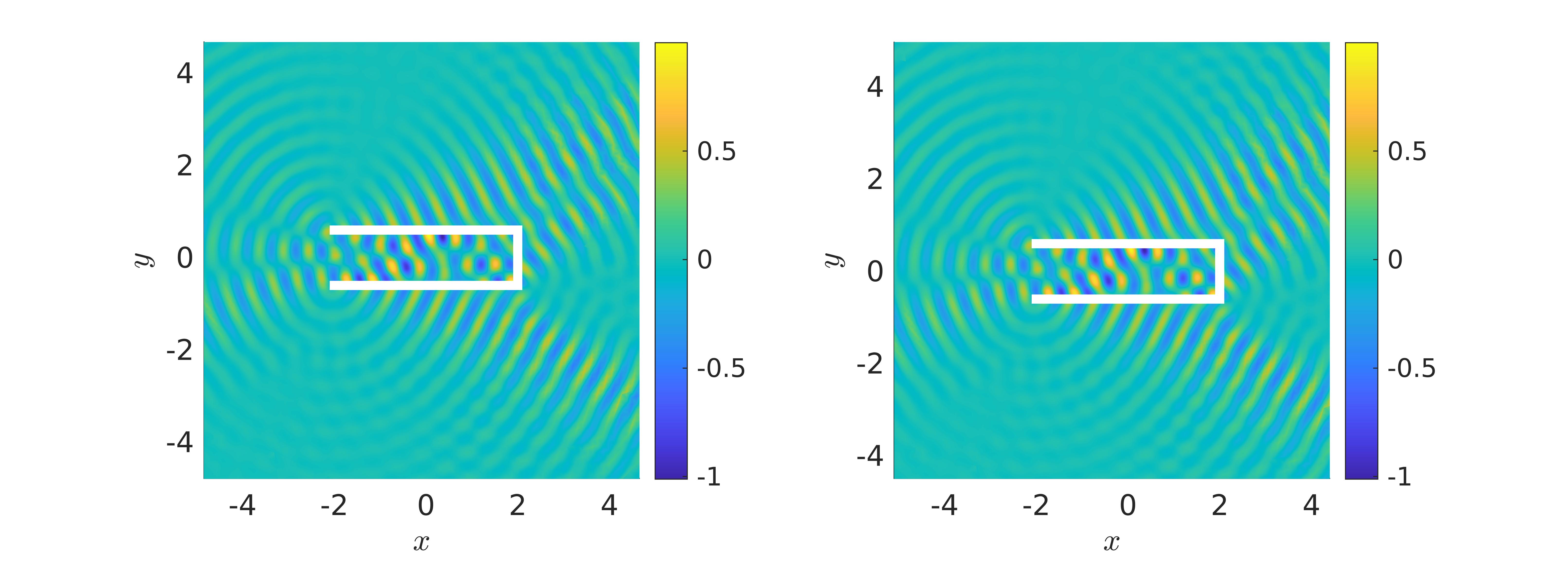}
\caption{Scattered $Hz$: uniform $p=8$ vs. adaptive $p=[1,\cdots,8]$}
\label{fig:contourHz_phi30}
\end{figure}

\begin{figure}
\centering
\includegraphics[width=0.8\textwidth]{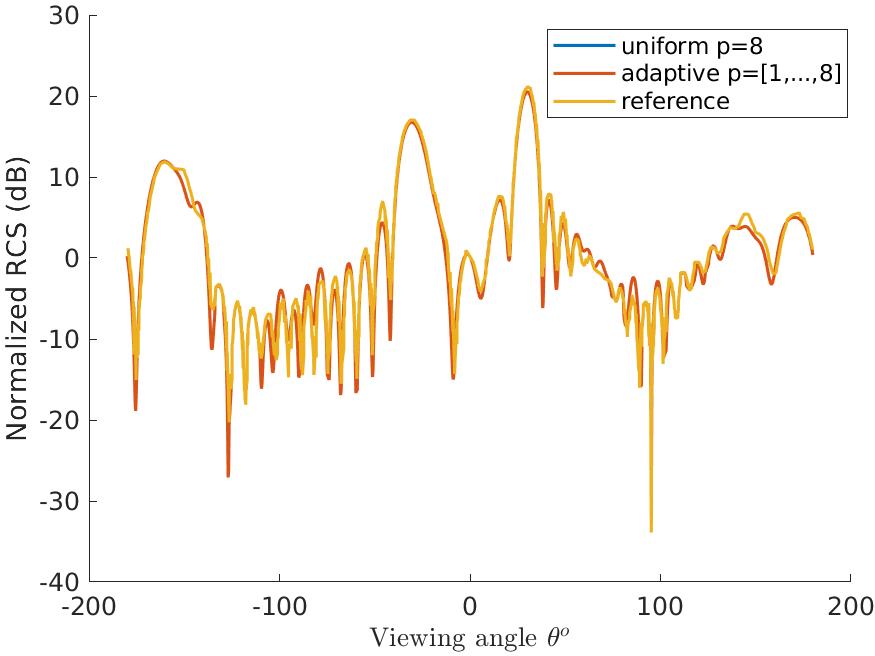}
\caption{RCS, semi-open cavity $8\lambda \times 2\lambda$, $\phi=30^0$}
\label{fig:rcsSOcavity_phi30}
\end{figure}

Next, we consider $\phi=30^0$. The $H_z$ plots for uniform $p=8$ and adaptive $p=[1,\cdots,8]$ are shown in fig. \ref{fig:contourHz_phi30} and the RCS obtained, in fig. \ref{fig:rcsSOcavity_phi30}. The simulation takes longer to converge due to multiple internal reflections and takes up to 50 periods. The reference solution is a 8th finite element frequency domain solution from \cite{Ledger2002}. There is good agreement between the adaptive and reference solutions except the sharpness of the troughs in a few places which is usually not expected to match exactly. Also noticeable, is that the plots for the adaptive and uniform cases are indistinguishably close, which is true of all results presented further as well. Since a variable $p$ method cannot be expected to perform any better than its uniform counterpart, it is reasonable to consider the adaptive algorithm effective.

\subsection{Two adjacent cylinders}
\begin{figure}
\centering
\scalebox{0.75}{
\begin{tikzpicture}[
photon/.style={decorate, decoration={snake}, draw=black}]
\node[] (orig) at (-0.5,4) {TE};
\node[] (phi) at (0.5,0.5) {$\phi$};
\node[] (l) at (0,-3) {$\bm{l=6.4\lambda},\hspace{0.1cm}\bm{2a=4\lambda}$};
\filldraw[color=green!60, fill=green!5, very thick](-3,0) circle (2);
\filldraw[color=green!60, fill=green!5, very thick](3,0) circle (2);
\draw[thick,->,photon] (0,5) -- (0,2.9);
\draw[thin, dashed] (-3,2.5) -- (-3,-2.5);
\draw[thin, dashed] (3,2.5) -- (3,-2.5);
\draw[thin, dashed] (0,2.5) -- (0,-2.5);
\draw[thin, dashed] (-1,0) -- (1,0);
\draw[thin, -Stealth] (.5,0) arc  (0:90:0.5);
\draw[thick, {Stealth[scale=1]}-{Stealth[scale=1]},blue] (-3,-1) --node [pos=0.45,above,black] {$\bm{l}$} (3,-1) ;
\draw[thick, {Stealth[scale=1]}-{Stealth[scale=1]},blue] (-5,0) --node [pos=0.4,above,black] {$\bm{2a}$} (-1,0);
\draw[thick, {Stealth[scale=1]}-{Stealth[scale=1]},blue] (1,0) --node [pos=0.4,above,black] {$\bm{2a}$} (5,0);
\end{tikzpicture}}
\caption{Schematic for the scattering off 2 adjacent cylinders problem}
\label{fig:schematic_2cyl}
\end{figure}
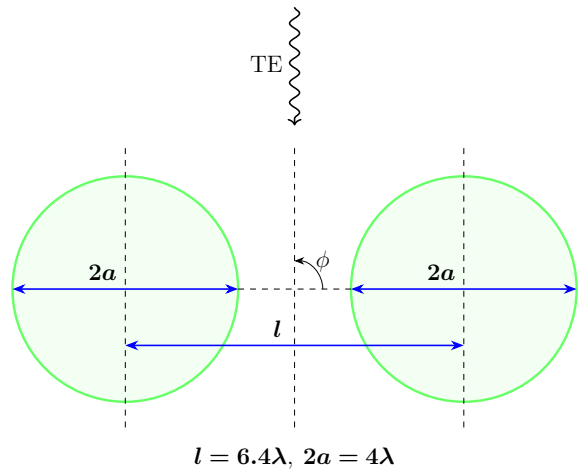

\par Another canonical problem to address the complexity of multiple reflections and reciprocal interactions is the that of multiple scatterers \cite{YoungBertrand}. In this section, we present results for the problem of two adjacent circular cylindrical scatterers, the schematic for which is shown in fig. \ref{fig:schematic_2cyl}. An incident TE wave illuminates the two cylinders in the configuration shown with the angle of incidence $\phi=270^0$. The scatterers are $6.4$ wavelengths apart and the diameter of each of them is 4 times the wavelength of the incident wave, thus lying in the optical scattering territory.

\begin{figure}
\centering
\includegraphics[width=\textwidth]{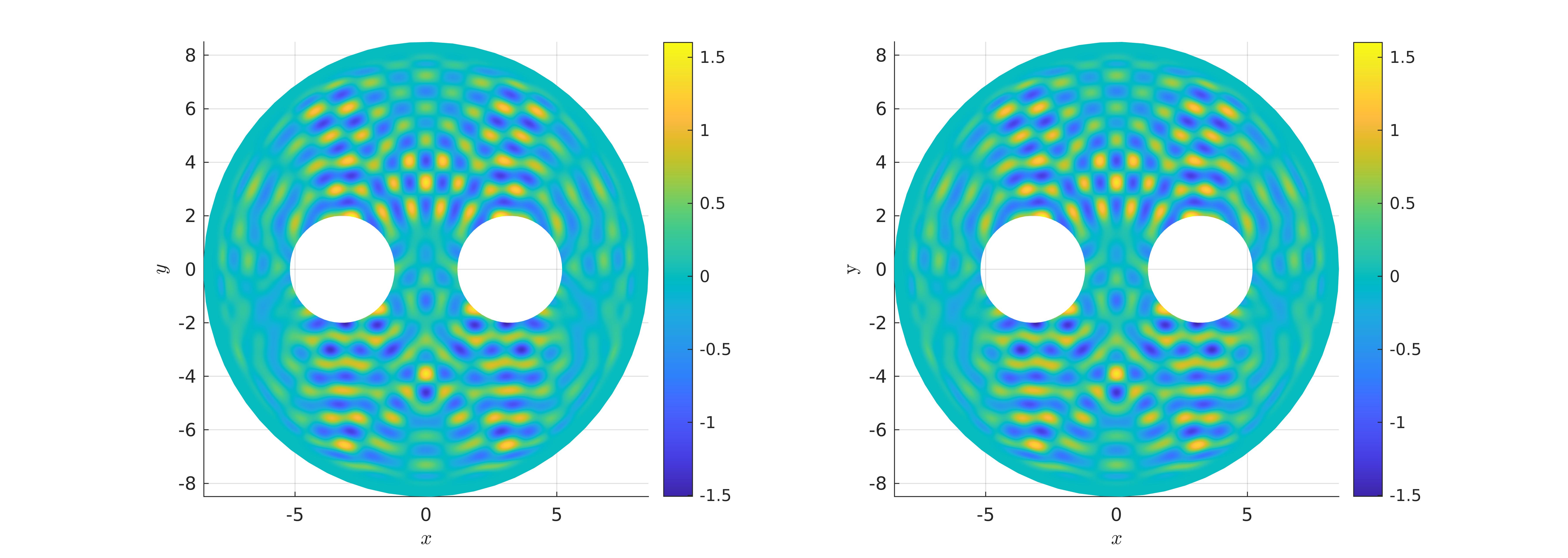}
\caption{Scattered $H_z$, 2 adjacent cylinders, $\phi=270^0$, uniform $p=8$ vs. adaptive $p=[1,\cdots,8]$}
\label{fig:2cyl_contourHz}
\end{figure}

\begin{figure}
\centering
\includegraphics[width=0.8\textwidth]{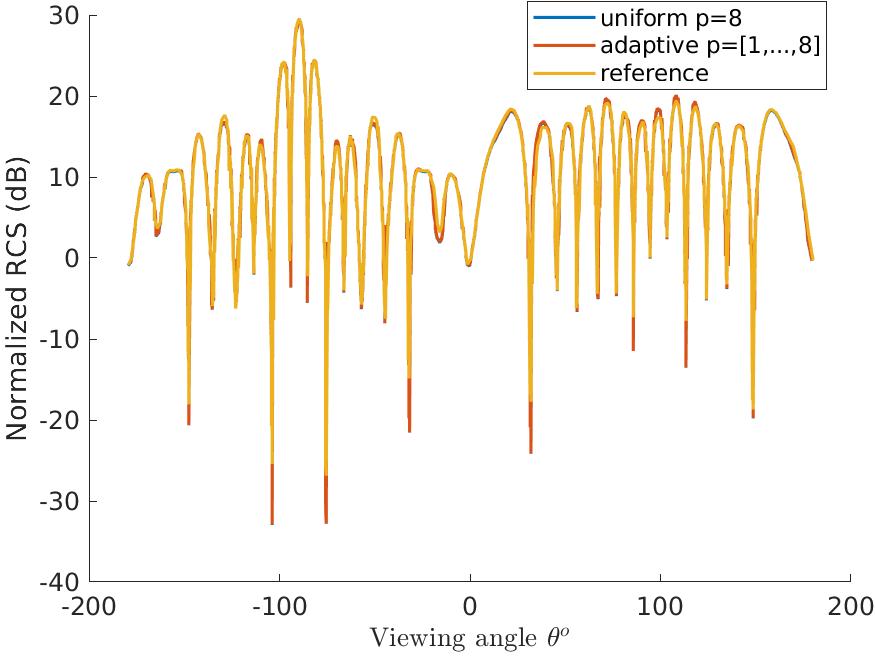}
\caption{RCS, 2 adjacent cylinders, $\phi=270^0$}
\label{fig:2cyl_rcs}
\end{figure}

Fig. \ref{fig:2cyl_contourHz} shows instantaneous scattered $H_z$ in harmonically steady state, comparing the uniform $p=8$ and adaptive $p=[1,\cdots,8]$ methods. They can be seen to be in accordance with each other. The corresponding RCS plots in fig. \ref{fig:2cyl_rcs} are compared with a 6th order accurate hybrid Galerkin solution \cite{Davies2009}. As in the case of the semi-open cavity , the plots corresponding to the uniform and the adaptive methods are close to each other and agree with that of the reference solution.

\subsection{3 adjacent cylinders}
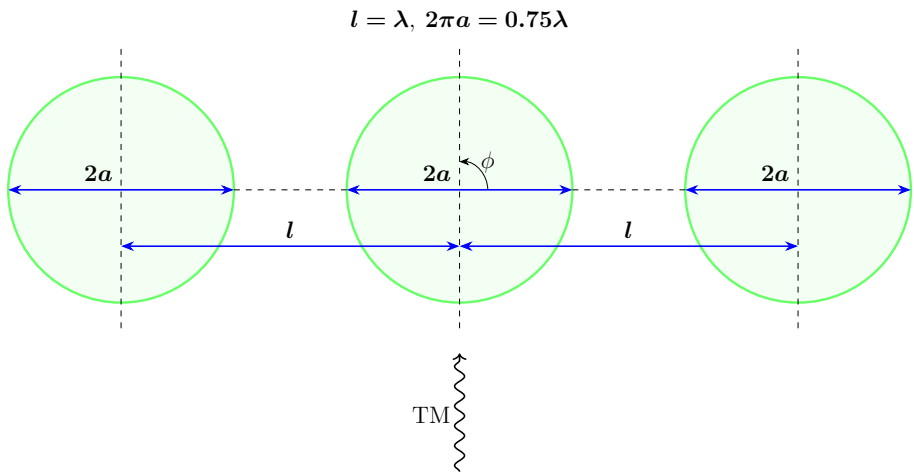
\begin{figure}
\centering
\scalebox{0.75}{
\begin{tikzpicture}[
photon/.style={decorate, decoration={snake}, draw=black}]
\filldraw[color=green!60, fill=green!5, very thick](0,0) circle (2);
\filldraw[color=green!60, fill=green!5, very thick](-6,0) circle (2);
\filldraw[color=green!60, fill=green!5, very thick](6,0) circle (2);
\node[] (phi) at (0.5,0.5) {$\phi$};
\node[] (orig) at (-0.5,-4) {TM};
\node[] (l) at (0,3) {$\bm{l=\lambda},\hspace{0.1cm} \bm{2\pi a=0.75\lambda}$};
\draw[thick,->,photon] (0,-5) -- (0,-2.9);
\draw[thin, dashed] (-6,2.5) -- (-6,-2.5);
\draw[thin, dashed] (0,2.5) -- (0,-2.5);
\draw[thin, dashed] (6,2.5) -- (6,-2.5);
\draw[thin, dashed] (-4,0) -- (-2,0);
\draw[thin, dashed] (2,0) -- (4,0);
\draw[thin, -Stealth] (.5,0) arc  (0:90:0.5);
\draw[thick, {Stealth}-{Stealth},blue] (-6,-1) --node [midway,above,black] {$\bm{l}$} (0,-1);
\draw[thick, {Stealth}-{Stealth},blue] (0,-1) --node [midway,above,black] {$\bm{l}$} (6,-1);
\draw[thick, {Stealth}-{Stealth},blue] (-8,0) --node [pos=0.4,above,black] {$\bm{2a}$} (-4,0);
\draw[thick, {Stealth}-{Stealth},blue] (4,0) --node [pos=0.4,above,black] {$\bm{2a}$} (8,0);
\draw[thick, {Stealth}-{Stealth},blue] (-2,0) --node [pos=0.4,above,black] {$\bm{2a}$} (2,0);
\end{tikzpicture}}
\caption{Schematic for the scattering off 3 adjacent cylinders problem}
\label{fig:schematic_3cyl}
\end{figure}

\par Continuing with the theme of numerous scatterers giving rise to multiple reflections off each other, the next illustration has three adjacent scatterers \cite{Ragheb1985}. A schematic of the problem is shown in fig. \ref{fig:schematic_3cyl}. An incident TM wave strikes at an angle of incidence $\phi=90^0$, scatterers each of circumference $0.75\lambda$ and placed one wavelength apart axially. Comparisons are made between the uniform $p=8$ and adaptive $p=[1,\cdots,8]$ methods.

\begin{figure}
\centering
\includegraphics[width=\textwidth]{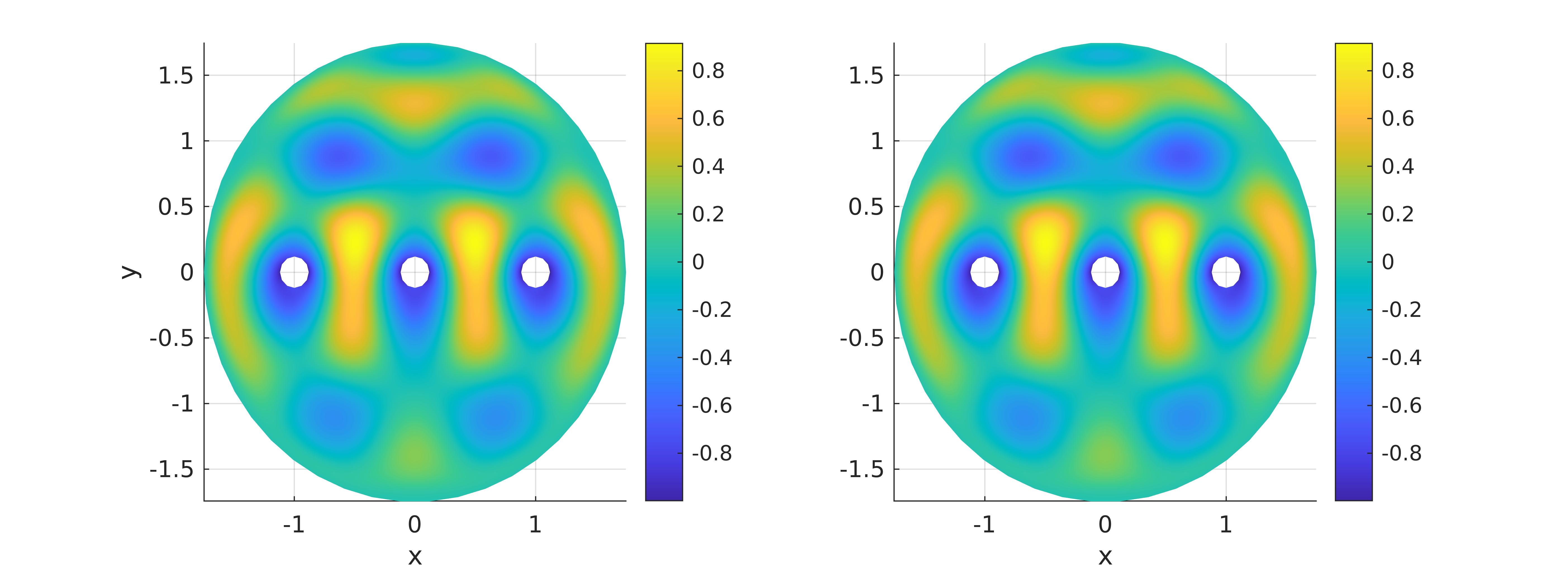}
\caption{Scattered $E_z$, 3 adjacent cylinders, $\phi=90^0$, uniform $p=8$ vs. adaptive $p=[1,\cdots,8]$}
\label{fig:3cyl_contourEz}
\end{figure}

\begin{figure}
\centering
\includegraphics[width=0.8\textwidth]{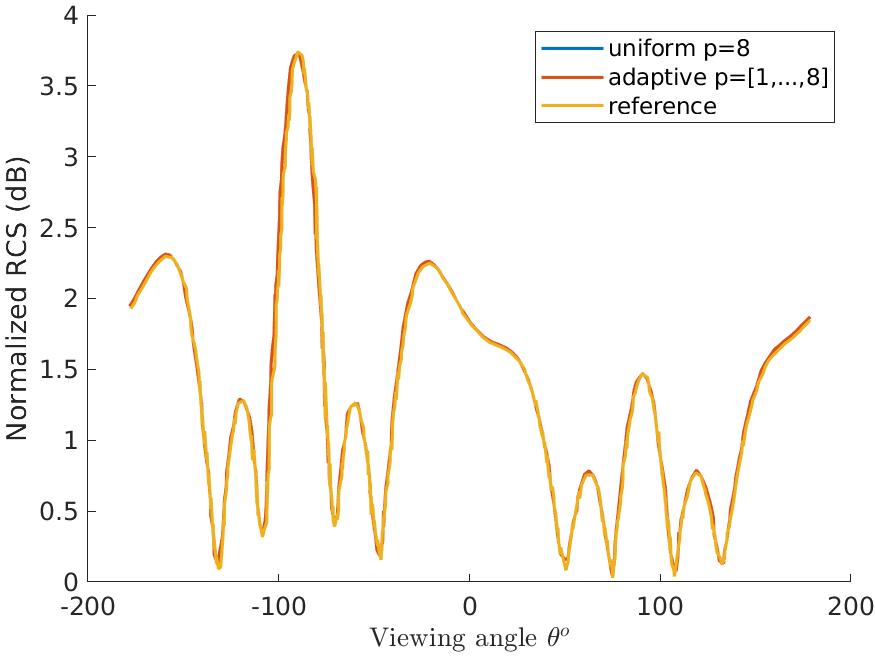}
\caption{RCS, 2 adjacent cylinders, $\phi=90^0$}
\label{fig:3cyl_rcs}
\end{figure}

The mutual interactions can be visualized using plots
of the instantaneous scattered $E_z$ in harmonically steady state, shown in fig. \ref{fig:3cyl_contourEz} and match with the ones in the reference solution \cite{Shah2021}. Fig. \ref{fig:3cyl_rcs} shows the RCS plots obtained with the uniform and adptive $p$ methods, compared to the reference solution which is obtained analytically \cite{Ragheb1985}. This set of test cases and the consistent agreement with the reference solutions, show the efficacy of the divergence driven $p$-adaptive algorithm proposed. 
\par The proposed algorithm serves to introduce the philosophy of utilizing numerical divergence in such adaptive methods, as opposed to eliminating it or leaving untreated. Divergence as a quantity can be exactly computed in real time, coupled with the fact that it does not require an estimation procedure, as with traditional truncation or discretization error based techniques \cite{Roy,Dolejsi2015,Rueda-Ramirez2019}.

\subsection{Computational performance}
\begin{table}
\begin{center}
\begin{tabular}{|c|c|c|}
\hline
\begin{tabular}{@{}c@{}}\textbf{Problem} \\ \textbf{size, illumination}\end{tabular} & \textbf{Savings in DOFs (\%)} & \textbf{effective $p$} \\
\hline
 2$\lambda$, TM & 43.14 & 5.31 \\ 
 2$\lambda$, TE & 43.41 & 5.30 \\
 15$\lambda$, TM & 34.36 & 5.90 \\
 15$\lambda$, TE & 35.01 & 5.87 \\ \hline
\end{tabular}
\end{center}
\caption{Computational savings in DOFs and effective $p$ employed to achieve accuracy corresponding to a uniform 8th order method.}
\label{table:comper}
\end{table}
In this section, we measure the computational gain obtained using the proposed divergence error driven $p$-adaptive method. The problem of scattering off a circular cylinder as shown in fig. \ref{fig:schematic} is used, with variations in electrical size 2$\lambda$, 15$\lambda$ and the incident illumination TM, TE. The 15$\lambda$ cases use a mesh two-thirds as dense as the one used in the 2$\lambda$ cases on a points per wavelength basis. We compare harmonically steady $p$-adaptive solutions with $p$ varying between $[1,\cdots,8]$, to a uniform $p=8$ solution in all cases and present the savings achieved in the time averaged number of DOFs required to achieve similar accuracy. Also, an average $p$, averaged over both time and space, is shown in table \ref{table:comper} to indicate the effective $p$ used to achieve $8$th order accurate solutions.

\begin{figure}[h!]
\centering
\begin{subfigure}{0.48\textwidth}
\includegraphics[width=\textwidth]{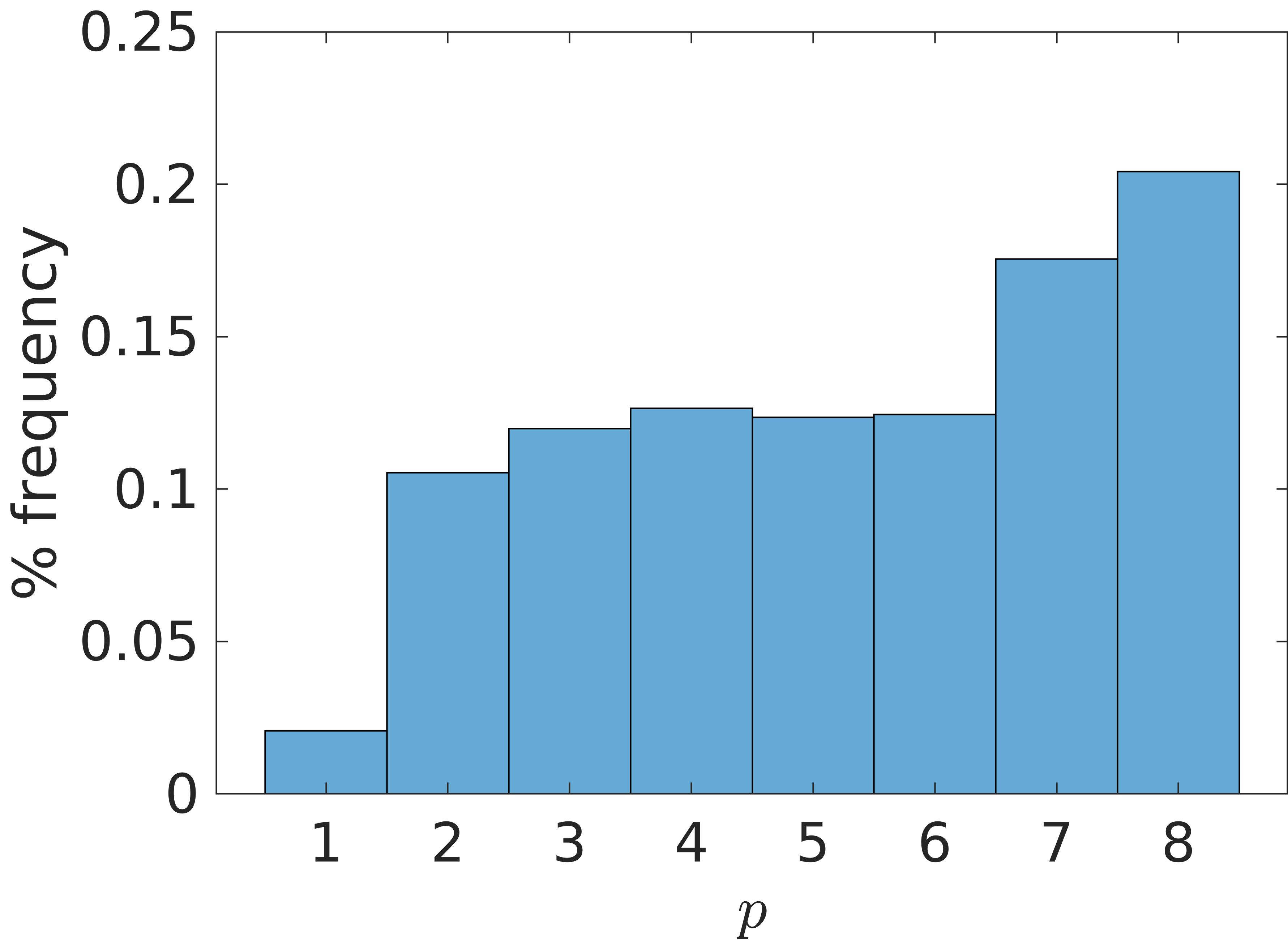}
\caption{2$\lambda$, TM }
\end{subfigure}
\hfill
\begin{subfigure}{0.48\textwidth}
\includegraphics[width=\textwidth]{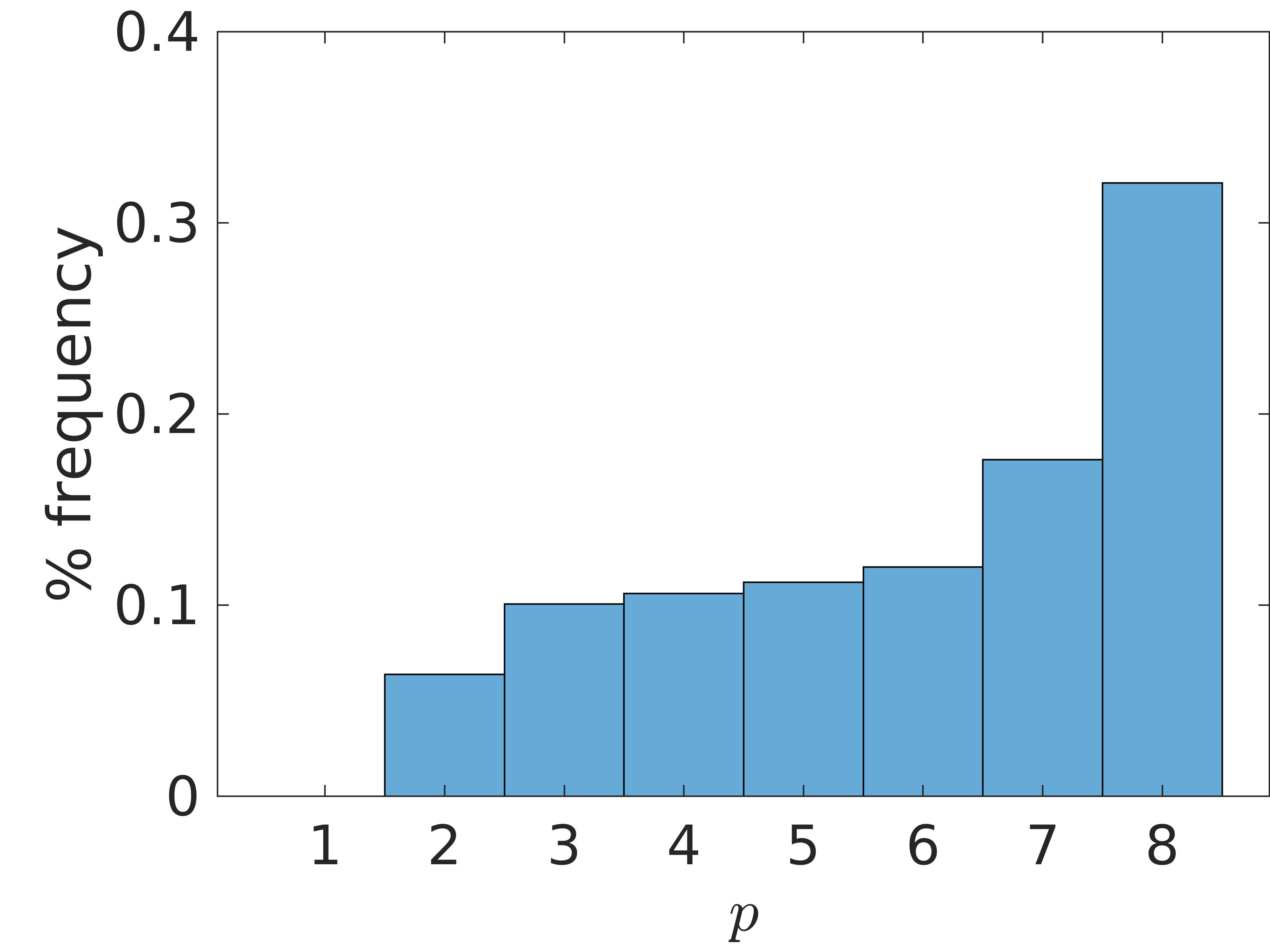}
\caption{15$\lambda$, TM}
\end{subfigure} \\
\begin{subfigure}{0.48\textwidth}
\includegraphics[width=\textwidth]{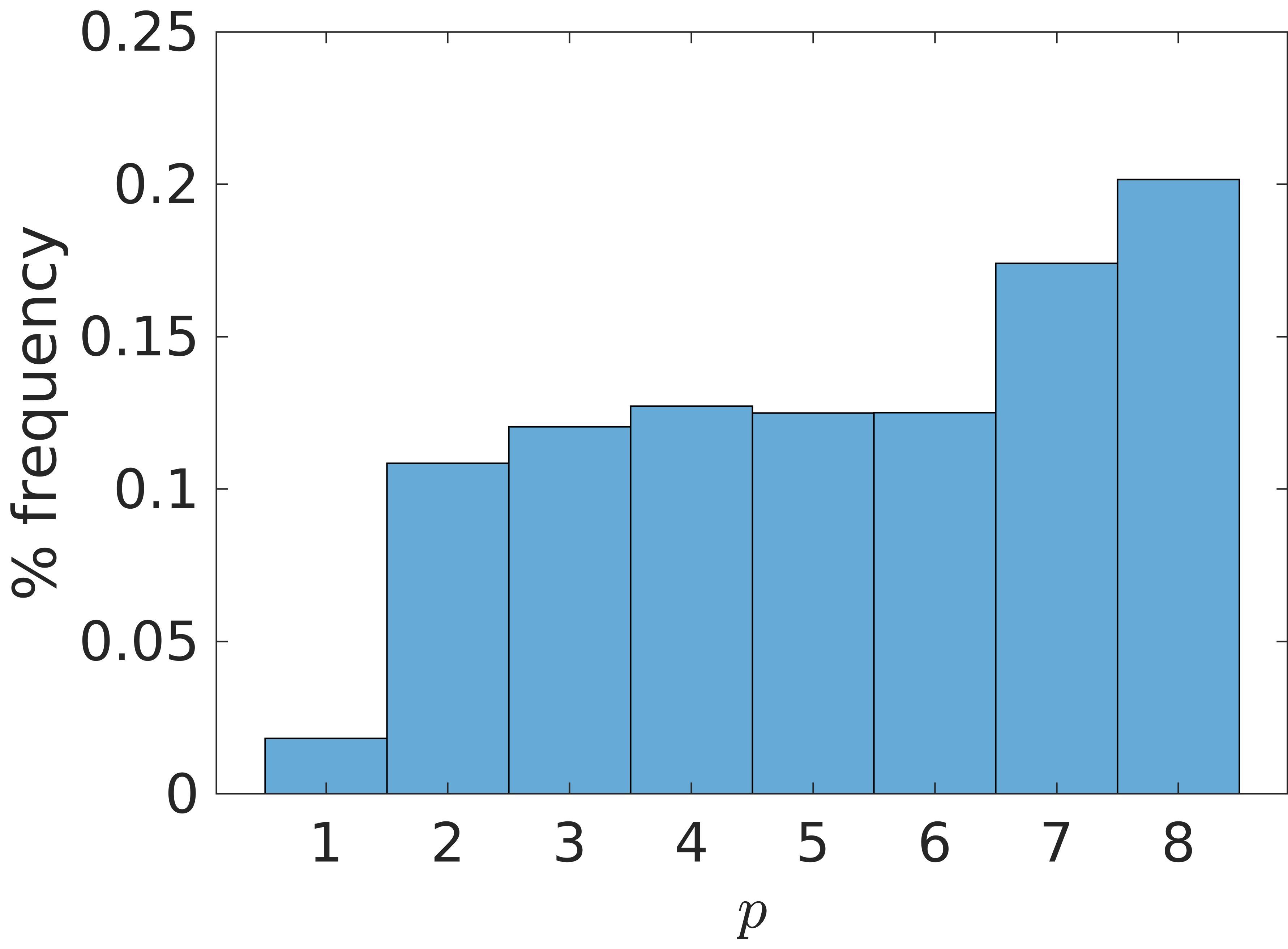}
\caption{2$\lambda$, TE}
\end{subfigure}
\hfill
\begin{subfigure}{0.48\textwidth}
\includegraphics[width=\textwidth]{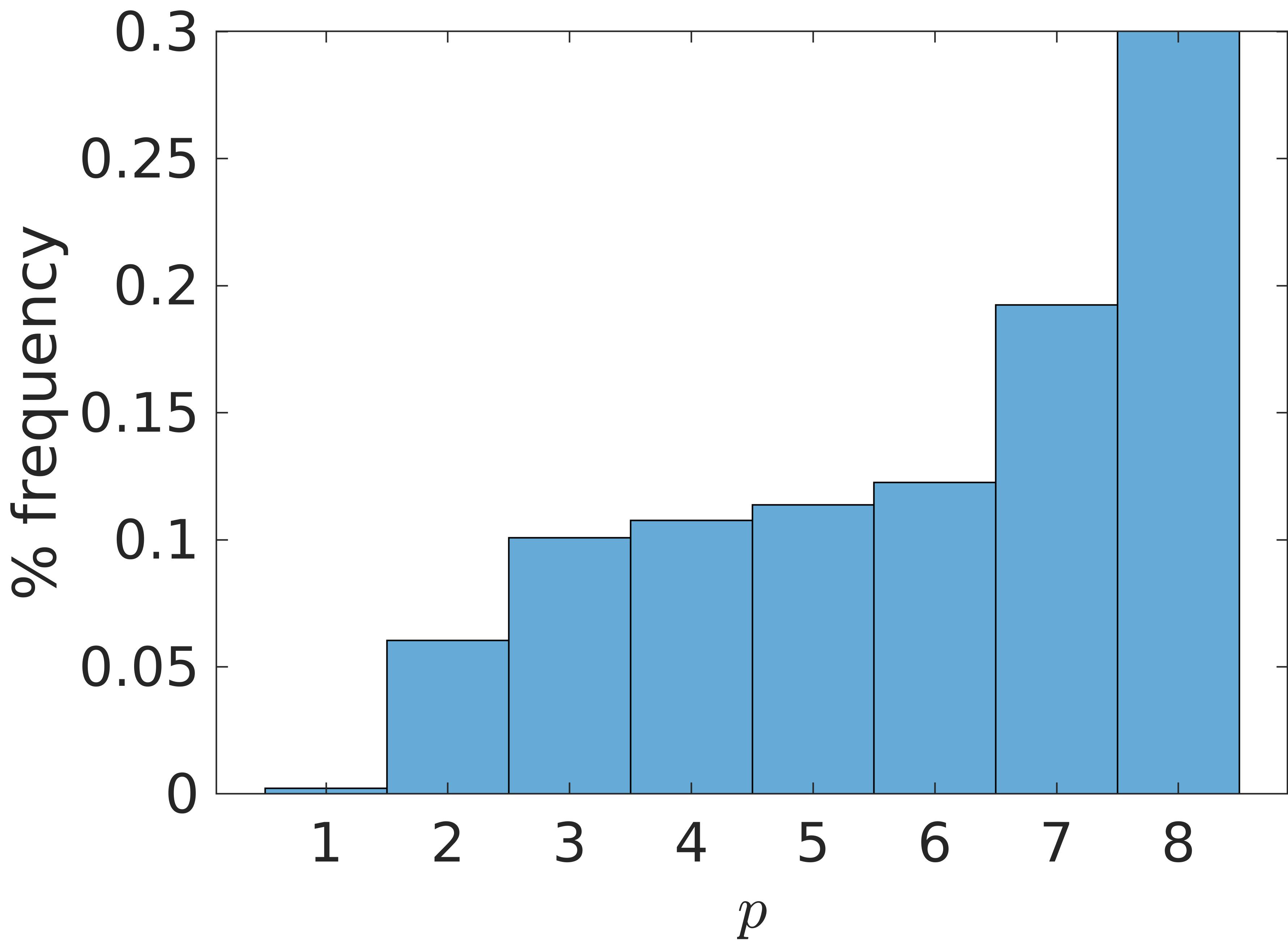}
\caption{15$\lambda$, TE}
\end{subfigure}
\caption{Histogram of time averaged $p$ for various scatterer sizes and incident illumination in the cylinder scattering problem.}
\label{fig:histogram}
\end{figure}

Fig. \ref{fig:histogram} shows histograms of time-averaged $p$ over the computational domain. A larger scatterer, here the 15$\lambda$ in size, naturally uses more $p=8$ cells while expectedly the incident illumination does not affect the distribution. The savings in DOFs show the cost effectiveness of the divergence error as driver for $p$-adaptive methods.

\section{Conclusion}
A divergence driven $p$-adaptive method is shown to work effectively in wave dominated problems of electromagnetic scattering in DGTD methods. The proposed $p$-variable method uses a combination of spatial operators of varying orders of accuracy across the computational domain, and is shown to be as good as a uniform order method that uses the highest level accuracy throughout, in terms of the resulting quantities of interest. A theory with regard to divergence error is formulated and fundamental relations between relative truncation and relative divergence errors have been established analytically. Illustrations using a simple $p$-adaptive algorithm based on the proposed theory have been presented to show the effectiveness of the readily computable divergence error, used as a driver in such algorithms.
\par Moreover, a novel perspective is presented in the context of treatment of the numerical divergence errors appearing in such simulations. The underutilized resource of divergence error is shown to have potential to drive adaptive algorithms as it acts as a proxy to truncation error. Traditionally, divergence errors when present, are eliminated by modelling the system to closely satisfy the divergence constraints or making dedicated effort to clean them explicitly. It is also common to not eliminate the divergence errors, where their presence may not deteriorate either the solution or the objective functionals. Extending this discussion, it is shown that there is an opportunity to utilize the divergence error to computational gain, at the algorithm layer. Divergence error based driver is a scalar and co-ordinate independent. Also, it is inexpensive to compute and can be easily accommodated in existing code, overcoming significant limitations in building practical and effective adaptive methods.

\printbibliography
\end{document}